\documentclass[11pt,a4paper]{article}
\usepackage{iftex}
\ifPDFTeX
  \usepackage[T1]{fontenc}
  \usepackage[utf8]{inputenc}
  \usepackage{lmodern}
  \usepackage{textcomp}
  \DeclareUnicodeCharacter{2212}{\textminus}
  \DeclareUnicodeCharacter{00B7}{\textperiodcentered}
\else
  \usepackage{fontspec}
\fi
\usepackage[margin=25mm]{geometry}
\usepackage{needspace}
\usepackage{placeins}
\usepackage{amsmath,amssymb}
\usepackage{graphicx}
\usepackage{array,booktabs,longtable}
\usepackage[font=small,labelfont=bf,labelsep=period,justification=raggedright,singlelinecheck=false]{caption}
\usepackage{pdflscape}
\usepackage[authoryear,round]{natbib}
\usepackage{url}
\usepackage[hidelinks]{hyperref}
\newcommand{\multiheader}[1]{#1}



\title{Do LLM-derived text features improve electricity price forecasts beyond
document activity? Evidence from Australia's National Electricity Market}
\author{Fanxing Ning\thanks{School of Electrical and Computer Engineering, The University of Sydney, Sydney, Australia. Email: fnin0084@uni.sydney.edu.au}
  \and Jianguo Zhu\thanks{School of Electrical and Computer Engineering, The University of Sydney, Sydney, Australia. Corresponding author. Email: jianguo.zhu@sydney.edu.au}}
\date{}
\begin{document}
\maketitle
\begin{abstract}
Large language models (LLMs) can turn market notices and news into
features for electricity price forecasting. Such features also record
when documents arrive and whether their extraction failed, so a gain
measured against a model without text can reflect document activity as
well as content. We add the counts of the same documents and of their
failed extractions to a no-text LightGBM model and measure each LLM
representation against this activity reference, which needs no
additional LLM requests. With gpt-6-luna as the extractor, we forecast
Australian electricity prices from operator notices (5--30 minutes
ahead) and WattClarity news (day-ahead) after the extractor's documented
knowledge cutoff, averaging forecasts over three training seeds. For
news, the counts alone lower mean absolute error (MAE) by 3.0\%, most of
the 3.6\% reduction that the better LLM representation achieves relative
to the no-text model; its increment beyond the counts is not
significant. For notices, two of the three LLM representations lower MAE
beyond the counts, by 0.14--0.21\% and 0.45--0.52\%, whether the counts
are cumulative or cover recent windows, whereas the third shows no significant
reduction. Time-shifted placebos give mixed evidence on whether these
gains require text aligned with the target. Benchmarks using the
operator's pre-dispatch forecasts (notices) or price history alone
(news) are more accurate than all these text models. Gains against the
no-text model and the activity reference differ in sign at individual
seeds and, for notices with cumulative counts, in seed-averaged
forecasts. Text gains should be reported against both references.
\end{abstract}
\section{Introduction}\label{introduction}

Australia's National Electricity Market (NEM) is a demanding setting for
electricity price forecasting \citep{Weron2014}: prices are volatile,
with recurrent spikes and heavy tails
\citep{Christensen2012Spikes,Cornell2024}. Large language models (LLMs)
can turn market notices and news into headline assessments or structured
fields for a forecasting model
\citep{LopezLiraTang2026,Chen2026REPrice}, and several recent studies
use LLMs to add notice or news information to NEM price forecasts
\citep{Liu2024bExtreme,Lu2025Agents,Bi2025NSWEPNews,Chen2026REPrice,Huang2025ShortTerm}.
These studies measure the gain against a model without text, but
features built from documents also record when documents arrive and
whether their extraction failed. Documents cluster on eventful days and
close to the events they describe, so their arrival alone can predict
prices. A gain beyond document activity therefore needs a reference
model that receives the same document and failure counts but no content.

This paper asks whether LLM-extracted content improves price forecasts
beyond document activity: the counts of documents available when each
forecast is issued and of those whose extraction failed. It makes two
contributions. First, it introduces a matched activity reference that
adds these counts to a no-text model; each LLM representation is then
added to the reference (Section~4, Figure~1). The reference uses the
same documents and extraction run, so it requires no additional LLM
requests, and it splits a gain over the no-text model into an activity
step and a representation step. Controls for document volume are
established in text-as-data research (Section~2), but none of the
electricity studies in Table~1 reports such a reference. Second, it
shows that the counts alone achieve most of the news gain over the
no-text model and that, in both tasks, the LLM representations add small
or no gains beyond them (Section~6).

\begin{figure}[!t]
\centering
\includegraphics[width=\textwidth]{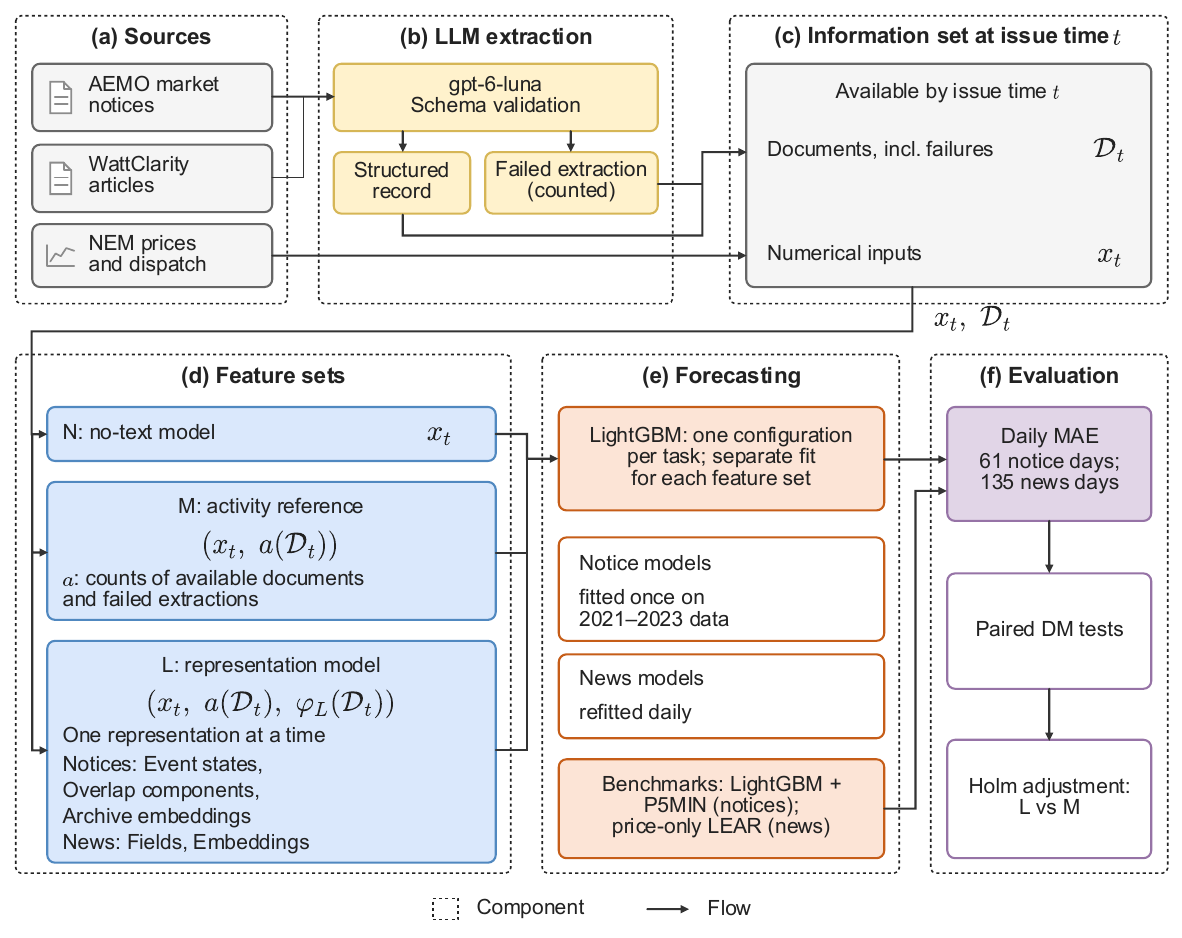}
\caption{Study design. (a) Sources. (b) Offline extraction yields one record per
saved document version; a version whose final extraction fails
validation is kept as a failed extraction. (c) Numerical inputs and
documents available at each issue time. (d) Feature sets of the no-text
model N, the activity reference M and a representation model L, which
adds one representation at a time. (e) Forecasting models. (f)
Evaluation. Not shown are the non-LLM representations, which read the
available source text, and the time-shifted placebos, which rebuild the
text inputs from the documents available 28 days earlier (Section~4.3).}\label{fig:1}
\end{figure}

The forecasts are evaluated after the LLM's documented knowledge cutoff,
which none of the LLM-based electricity studies in Table~1 reports. This
limits look-ahead, since an LLM whose training data overlap the
evaluation period may carry knowledge of later outcomes into its
features \citep{GlassermanLin2023,SarkarVafa2024,LopezLiraTangZhu2025}.
Time-shifted placebos, which rebuild the same text inputs from the
documents available 28 days earlier, test whether a gain requires text
aligned with the target (Section~4.3).

Market participants follow the notices in which the Australian Energy
Market Operator (AEMO) announces reserve conditions and interventions,
and they also follow commentary and news, such as WattClarity's reports
on NEM events. We use both in two forecasting tasks: AEMO market notices
for five-minute prices 5 to 30 minutes ahead in all five NEM regions,
and WattClarity articles for day-ahead forecasts, issued at midnight, of
the 24 hourly prices in New South Wales, Queensland and Tasmania. The
extractor, gpt-6-luna, reads one document at a time and returns
schema-validated fields. From these we build three LLM representations
of the notices and two of the articles. For notices, Event states
summarise the events relevant to the target interval, Overlap components
count the messages whose windows overlap it, and the Archive embeddings
representation encodes all extracted records to date. For articles, the
Fields representation counts the extracted records by category, and the
Embeddings representation encodes them. The non-LLM representations are
rule-parser records, word frequencies and raw-text embeddings of the
same documents (Section~4). Figure~1 outlines the design.

The notice window covers the 61 days of August and September 2026, and
the news window the 135 days from 19 May to 30 September 2026. Each task
uses one fixed LightGBM configuration, and the no-text model, the
activity references and the LLM representation models are fitted at
three training seeds. The notice benchmark, LightGBM + P5MIN, adds
inputs from AEMO's five-minute pre-dispatch forecast (P5MIN) to the
no-text model; the news benchmark is LEAR, a price-only autoregressive
model estimated with the least absolute shrinkage and selection operator
(LASSO).

The document counts alone achieve most of the news gain. With forecasts
averaged over the three seeds, adding the article and failed-extraction
counts to the no-text model lowers the mean absolute error (MAE) by
3.03\%. The Embeddings model lowers MAE by 3.55\% relative to the
no-text model and by 0.53\% relative to the activity reference, an 
insignificant improvement. The Fields model raises MAE (Section~6.1).

Beyond the activity features, two of the three LLM notice
representations lower MAE slightly. With seed-averaged forecasts, Event
states and Archive embeddings lower MAE by 0.14\% and 0.52\% relative to
the activity reference, and Overlap components does not significantly
lower it. These steps are similar when the two cumulative counts of the
notice reference are replaced by counts over the past seven days or 24
hours (Sections~4.2 and 6.1). At a single seed, the Archive embeddings
model is more accurate than its time-shifted placebo with the original
reference but not with the seven-day reference. At that seed,
rule-parser and word-frequency representations of the notices give
reductions within the range of the LLM representations (Section~6.2).

Gains measured against the activity reference and against the no-text
model differ in sign at individual seeds and, with the original
cumulative notice counts, also for Event states and Overlap components
in the seed-averaged forecasts (Section~6.1). Both task benchmarks
remain more accurate than all text models built on the no-text model.

The rest of the paper is organised as follows. Section~2 reviews the
related literature, and Section~3 describes the data, the forecasting
tasks and when each input becomes available. Section~4 describes the
extraction, the activity features and the text representations, and
Section~5 sets out the models and tests. Section~6 reports the results,
Section~7 discusses them, and Section~8 concludes. Appendices~A--D cover
alternative tests and intervals, information timing and learner
settings, the extraction and human review, and additional results. A
file in the code repository, cited as Supplement~A.1--D.2, gives further
implementation details.

\begin{table}[!t]
\begingroup\scriptsize\fontsize{8}{10.5}\selectfont
\setlength{\tabcolsep}{4pt}
\renewcommand{\arraystretch}{1.04}
\hyphenpenalty=10000\exhyphenpenalty=10000
\caption{Closest studies. Activity reference: a reference model that adds only
counts of the same documents and of their failed extractions; unmatched:
counts used as features but not as such a reference; timing control: how
the information available at each issue time is controlled; post-cutoff
evaluation: evaluation after the LLM's documented knowledge cutoff.
n.r.: not reported; n.a.: not applicable.}\label{tab:1}
\begin{tabular}{@{}>{\raggedright\arraybackslash}p{\dimexpr 0.12747120182210828\linewidth-1.75\tabcolsep\relax}>{\raggedright\arraybackslash}p{\dimexpr 0.12088133029648623\linewidth-1.75\tabcolsep\relax}>{\raggedright\arraybackslash}p{\dimexpr 0.14284756871522633\linewidth-1.75\tabcolsep\relax}>{\raggedright\arraybackslash}p{\dimexpr 0.10989821108711617\linewidth-1.75\tabcolsep\relax}>{\raggedright\arraybackslash}p{\dimexpr 0.1384543210314783\linewidth-1.75\tabcolsep\relax}>{\raggedright\arraybackslash}p{\dimexpr 0.1164880826127382\linewidth-1.75\tabcolsep\relax}>{\raggedright\arraybackslash}p{\dimexpr 0.1318644495058563\linewidth-1.75\tabcolsep\relax}>{\raggedright\arraybackslash}p{\dimexpr 0.1120948349289902\linewidth-1.75\tabcolsep\relax}@{}}
\toprule
Study & Market; target & Text; LLM output & Activity reference & Timing control & Post-cutoff evaluation & Benchmark & Statistical test \\
\midrule
\citet{Chen2026REPrice}, RE-Price & NEM; day-ahead hourly, probabilistic & WattClarity; impact tiers, attributes & None & Latest publication or edit time & n.r. & Same framework without news & n.r. \\
\addlinespace[1.5pt]
\citet{Huang2025ShortTerm} & NSW; five-minute & WattClarity; event labels & None & Daily protocol; news lagged one day & n.r. & XGBoost without news & DM test\(^{a}\) \\
\addlinespace[1.5pt]
\citet{Lu2025Agents} & NEM; next-day five-minute & ABC News, news.com.au; two agents & None & Past 20 days of news & n.r. & Without sentiment agent; AEMO pre-dispatch & n.r. \\
\addlinespace[1.5pt]
\citet{Bi2025NSWEPNews}, NSW-EPNews & NSW; next-day half-hourly & WattClarity summaries; LLM forecasts & None & Same-day prices and news & n.r. & Price-only ablation & n.r. \\
\addlinespace[1.5pt]
\citet{Liu2024bExtreme} & NEM; extreme prices & AEMO notices; event series & None & n.r. & n.r. & Own forecast without notice adjustment & n.r. \\
\addlinespace[1.5pt]
\citet{WangYu2025} & CAISO; 96--192~h & News; impact sentiment & None & From the day after publication & n.r. & TimeXer without news & n.r. \\
\addlinespace[1.5pt]
\citet{Yi2026TRACE}, TRACE & PJM; day-ahead hourly, probabilistic & PJM emergency procedures; LLM risk assessments & None & Published and not cancelled by the forecast cutoff & n.r. & Same models with text zeroed & Permutation test\(^{b}\) \\
\addlinespace[1.5pt]
\citet{Zhou2026Weather} & Solar output; real-time prices & Forecast discussions; weather-risk features & None\(^{c}\) & By issue date; lagged one day & n.r. & No-text and keyword-rule models & Daily-loss tests\(^{d}\) \\
\addlinespace[1.5pt]
\citet{Bai2024NewsLoad} & Day-ahead demand & News; no LLM & Counts, unmatched\(^{e}\) & Daily text, day \(d-1\) for day \(d+1\) & n.a. (no LLM) & Model without text & DM test \\
\addlinespace[1.5pt]
\citet{LopezLiraTang2026} & Stock returns & Headlines; one assessment each & None & Initial reaction separated from later drift & Yes & Simpler text representations & Regression t-tests\(^{f}\) \\
\addlinespace[1.5pt]
\citet{Kodeih2026} & USD/CAD; 1--24~h & Monetary-policy news; stance labels & Group ablation\(^{g}\) & Article timestamps aggregated by hour & n.r. & Random walk; group ablations & Bootstrap intervals\(^{h}\) \\
\addlinespace[1.5pt]
\textbf{This paper} & NEM; 5--30-min and day-ahead hourly & AEMO notices, WattClarity; one-document extraction & Matched counts\(^{i}\) & Proxy clock (notices); saved versions (news) & Yes (documented cutoff) & No-text model; activity reference; benchmarks & DM tests\(^{j}\) \\*
\bottomrule
\multicolumn{8}{p{\dimexpr\linewidth-2\tabcolsep\relax}}{\footnotesize\raggedright\setlength{\parindent}{0pt} \emph{Notes.} \(^{a}\) Overall gain not significant; significant in
spike periods. \(^{b}\) Correctly paired versus randomly re-paired text;
gains over the models with text zeroed are reported without a test.
\(^{c}\) Product counts are summed in the extraction cache, but the
forecasting roles use weather-risk scores. \(^{d}\) Paired daily losses
with a heteroskedasticity- and autocorrelation-consistent (HAC) variance
with seven lags, and seven-day block-bootstrap intervals. \(^{e}\)
Count-feature experiments. \(^{f}\) Standard errors clustered by firm
and date. \(^{g}\) Article counts and news-hour indicators form feature
groups; each group, including the LLM sentiment, is removed from the
full feature set in turn. \(^{h}\) With Benjamini--Hochberg correction.
\(^{i}\) Counts of the same available documents and of failed
extractions. \(^{j}\) Daily, with Holm adjustment per task; HAC (seven
lags) and Student-t sensitivities; block bootstrap.\par}\\
\end{tabular}\endgroup
\end{table}

\section{Related literature}\label{related-literature}

Media content predicts stock-market activity
\citep{Tetlock2007Incomplete}, and news text improves forecasts of
macroeconomic aggregates and commodity prices
\citep{LarsenThorsrud2019,Li2019,Kalamara2022,Ellingsen2022,Barbaglia2023,Abeyie2026}.
As in these studies, our text measures use only documents available at
the issue time \citep{Ellingsen2022,Kalamara2022,Barbaglia2023}, and
text is judged against benchmarks that already contain the relevant hard
data \citep{Bai2022CrudeOil}, with formal tests of predictive accuracy
\citep{Ardia2019,Kalamara2022,Barbaglia2023}. With LLMs, extraction must
also be compared with simpler representations of the same text.
\citet{LopezLiraTang2026} show that GPT-4 headline assessments capture
the initial market reaction and predict the subsequent price drift. We
adopt their comparison with simpler textual representations and evaluate
forecast targets after the extractor's documented knowledge cutoff.

LLMs enter time-series forecasting in two roles. As the forecaster, a
frozen LLM is reprogrammed for numerical series in Time-LLM
\citep{Jin2024TimeLLM}, and RIS-LLM forecasts electricity prices with an
LLM backbone and semantic labels derived from numerical features
\citep{Huang2026RISLLM}. As a reader, the LLM supplies information
absent from the numerical history: an agent selects and reasons over
news before forecasting \citep{Wang2024News}, Time-MMD pairs series with
sourced textual facts \citep{Liu2024aTimeMMD}, and Context is Key tests
whether forecasters use text that complements the history
\citep{Williams2025Context}. \citet{Majumder2024} and
\citet{Zhang2026Energy} discuss the capabilities and limitations of LLMs
in the energy sector.

Most LLM-based studies of NEM prices use the LLM as a reader, of either
operator notices or news. \citet{Liu2024bExtreme} fine-tune an open LLM
to turn AEMO market notices into an event series for extreme-price
forecasting. The closest design is RE-Price \citep{Chen2026REPrice},
which classifies the impact of WattClarity articles and extracts their
attributes for probabilistic day-ahead forecasts; both forecasting tasks
here adopt its document-level extraction approach (Section~4.1). From
the same source, \citet{Huang2025ShortTerm} add LLM event labels to
five-minute forecasts for New South Wales. \citet{Lu2025Agents} combine
a market-sentiment agent with a bidding-behaviour agent, and
\citet{Bi2025NSWEPNews} build a news-augmented benchmark and report
marginal or mixed news gains.

Related applications examine LLM news sentiment for prices in the
California Independent System Operator (CAISO) market
\citep{WangYu2025}, constrained weather-risk features from public
forecast discussions for solar and real-time-price forecasting
\citep{Zhou2026Weather}, and word frequencies, sentiment, topics and
embeddings for day-ahead electricity demand \citep{Bai2024NewsLoad}.

Text-as-data research treats text volume as a separate variable. The
number of news announcements is related to market activity
\citep{MitchellMulherin1994}. The economic policy uncertainty index is a
normalised count of articles on policy-related economic uncertainty
\citep{Baker2016EPU}, and \citet{Gentzkow2019Text} review such
text-based measures. Forecasting studies also separate volume from tone.
In volatility forecasts, the number of overnight firm news items helps
where firm-level sentiment adds little \citep{BodilsenLunde2025}. For
commodities, daily news volume predicts the direction of oil-price
volatility better than a heterogeneous autoregressive (HAR) benchmark
\citep{HashamiMaldonado2025}, whereas regressions on language-model text
representations outperform news-count benchmarks for agricultural
volatility, especially at longer horizons \citep{ParviniAssa2025}.
\citet{Adammer2025} construct news-coverage and sentiment predictors of
macroeconomic tail risk, and \citet{Kodeih2026} attribute exchange-rate
forecast gains to the timing, activity and LLM-derived sentiment of
policy communication.

Other designs test whether a gain depends on the text at all. In machine
learning, a partial-input baseline, such as a classifier given the
hypothesis without the premise, shows how much of a model's success
comes from shortcuts
\citep{Gururangan2018Artifacts,Geirhos2020Shortcut}. In time-series
forecasting, replacing the text given to three multimodal forecasters on
the Time-MMD benchmark with empty or shuffled text leaves their mean
squared error almost unchanged \citep{Sridhar2026}. A concurrent
benchmark pairs prices in the PJM Interconnection, a US electricity
market, with official operational text reconstructed as available at
each forecast cutoff; the text lowers upper-tail pinball loss, whereas
text mismatched across days raises it above the no-text baseline
\citep{Yi2026TRACE}. The activity reference applies the partial-input
logic to document-derived features, with counts taken from the same
documents and extraction run as the representation.

A text gain is credible only if the text was available when the forecast
was made: leakage of information unavailable at the issue time inflates
backtests \citep{Kaufman2011Alternative,KapoorNarayanan2023}. LLMs can
also introduce look-ahead through pretraining on text written after the
issue time
\citep{GlassermanLin2023,SarkarVafa2024,LopezLiraTangZhu2025}, and
\citet{Ludwig2024Econometric} show that valid predictions from text
require no leakage between the LLM's training data and the research
sample. One remedy is to test only after the model's documented
knowledge cutoff \citep{LopezLiraTang2026}; another is to use models
trained only on earlier text \citep{Drinkall2024,He2025}.

Gains from complex models often shrink against strong benchmarks and
ablations. For electricity prices, \citet{Lago2021} recommend strong
open benchmarks such as LEAR and paired Diebold--Mariano (DM) tests of
predictive accuracy. In the NEM, \citet{Cornell2024} compare their
forecasts with AEMO's pre-dispatch prices, and newer deep time-series
models do not always outperform standard deep-learning baselines
\citep{Gani2026}. \citet{Tan2024} report that removing the LLM from
three LLM-based time-series forecasters, or replacing it with a simple
attention layer, usually leaves accuracy unchanged or improves it. Of
the five NEM studies in Table~1, only \citet{Huang2025ShortTerm} report
a formal test of predictive accuracy.

\FloatBarrier

\section{Data}\label{data}

\subsection{Prices and the operator
forecast}\label{prices-and-the-operator-forecast}

The NEM spans five regions: New South Wales (NSW1), Queensland (QLD1),
South Australia (SA1), Tasmania (TAS1) and Victoria (VIC1). A regional
price is set for each five-minute dispatch interval, and generators can
rebid until dispatch \citep{Cornell2024}. Five-minute settlement
replaced 30-minute settlement in October 2021
\citep{CsereklyeiKhezr2024}, and our models train on data from that
month onward.

Targets and price history use AEMO's five-minute regional reference
prices from the first dispatch run, with intervention records removed.
The other numerical inputs come from AEMO's regional dispatch summaries
(Table~2). We use the final stored values because revision histories are
not available. All times use Australian Eastern Standard Time (AEST,
UTC+10), without daylight-saving shifts.

At short horizons, the operator publishes its own price forecast, P5MIN,
every five minutes. We use its records for the notice training sample
(October 2021--December 2023) and for 27 July--30 September 2026
(Supplement~B.1); it enters only the LightGBM + P5MIN benchmark and its
extension (Section~5.1).

\begin{table}[!hbp]
\begingroup\footnotesize
\setlength{\tabcolsep}{3pt}
\renewcommand{\arraystretch}{1.04}
\caption{Data sources and samples.}\label{tab:2}
\begin{tabular}{>{\raggedright\arraybackslash}p{\dimexpr 0.2\linewidth-2\tabcolsep\relax}>{\raggedright\arraybackslash}p{\dimexpr 0.21\linewidth-2\tabcolsep\relax}>{\raggedright\arraybackslash}p{\dimexpr 0.22\linewidth-2\tabcolsep\relax}>{\raggedright\arraybackslash}p{\dimexpr 0.235\linewidth-2\tabcolsep\relax}>{\raggedright\arraybackslash}p{\dimexpr 0.135\linewidth-2\tabcolsep\relax}}
\toprule
Source or sample & Content & Period & Size & Use \\
\midrule
AEMO dispatch prices & Five-minute regional reference prices, first dispatch run & Oct 2021--Sep 2026 & Continuous & Targets; price history; anchor \\
\addlinespace[0.35ex]
AEMO regional dispatch summaries & Demand; available and dispatchable generation; net interchange; demand
forecasts & Oct 2021--Sep 2026 & Continuous & No-text model inputs \\
\addlinespace[0.35ex]
AEMO P5MIN & Five-minute pre-dispatch price forecasts & Oct 2021--Dec 2023; 27 Jul--30 Sep 2026 & Continuous & LightGBM + P5MIN and its extension \\
\addlinespace[0.35ex]
AEMO market notices & Saved notices with source clocks & Oct 2021--Sep 2026; gaps Aug 2024--Oct 2025 and 19 May--26 Jul 2026 & 21,470; training sample 11,348; evaluation period 796 & Notice text features \\
\addlinespace[0.35ex]
WattClarity articles & Saved body with publication and modification times & Sep 2021--Sep 2026 & 2,506 articles & News text features \\
\addlinespace[0.35ex]
\addlinespace[0.6ex]
\multicolumn{5}{l}{\emph{Evaluation samples}}\\*
Notices & Horizons 5--30 min; five regions & Target days 1 Aug--30 Sep 2026 & 61 days; 527,040 forecasts & Notice comparisons \\
\addlinespace[0.35ex]
News & 24 hourly targets; NSW1, QLD1, TAS1 & Issue days 19 May--30 Sep 2026 & 135 days; 9,720 forecasts; 207 articles & News comparisons \\
\addlinespace[0.35ex]
News, extended window & As above & Issue days 1 Mar--30 Sep 2026 & 214 days; 15,408 forecasts & Sensitivity only (Table~D1) \\
\bottomrule
\end{tabular}\endgroup
\end{table}

\subsection{Notices and news}\label{notices-and-news}

AEMO market notices are the operator's messages to participants. Some
forecast a lack of reserve in a region, with the affected interval and
the reserve figures; others update or cancel earlier notices.
WattClarity, the news source of the closest price-forecasting studies
\citep{Bi2025NSWEPNews,Chen2026REPrice,Huang2025ShortTerm}, publishes
articles that report and analyse NEM events.

The saved notice archive has two gaps, August 2024--October 2025 and 19
May--26 July 2026 (inclusive), which are coded as missing data. Notices
in the training sample carry AEMO's official timestamps. The notices
saved for the evaluation period, 27 July--30 September 2026, start five
days before the notice window. Each article is stored as one saved body
with its publication and modification times. Earlier versions of an
edited article cannot be recovered. Table~2 lists the sources and
archive sizes; Appendix~C.1 reports the extraction runs and failures.

\subsection{Forecasting tasks, windows and information
sets}\label{forecasting-tasks-windows-and-information-sets}

Notice forecasts are issued every five minutes for the five-minute
prices 5, 10, 15, 20, 25 and 30 minutes ahead in all five regions, which
gives 8,640 forecasts per target day. News forecasts are issued at 00:00
AEST for the 24 hourly prices from 00:00--01:00 to 23:00--24:00 in NSW1,
QLD1 and TAS1, with lead times from zero to 23 hours, giving 72
forecasts per day. The NEM has no day-ahead market; we call these
forecasts day-ahead because they cover the 24 hours after their issue.
Each hourly target is the mean of the twelve five-minute prices in that
hour and is computed only when all twelve are present, without
interpolation, clipping or transformation.

OpenAI documents a knowledge cutoff of 18 May 2026 for gpt-6-luna
\citep{OpenAI2026Luna}, and this cutoff sets the evaluation windows
(Figure~2). The notice window, 1 August--30 September 2026 (61 target
days, 527,040 forecasts), lies wholly after this cutoff and begins after
the second archive gap; notices from 27--31 July enter the information
set. To limit look-ahead through pretraining (Section~2), the news
window covers issue days from 19 May to 30 September 2026: 135 days and
9,720 forecasts, with 207 distinct articles available. Every forecast
target falls after the cutoff; available articles can predate it through
the 72-hour lookback or later saved versions of older articles (Appendix~B.2). Article counts vary over the news window (Figure~2(d)), whereas
notice arrivals cluster on a few days (Figure~2(e)).

A numerical record enters a forecast once it is available under the
timing rule for its source, and the latest available price is the anchor
(Appendix~B.1). Notices enter five minutes after their timestamp,
rounded up to the five-minute grid. For the evaluation-period notices,
the timestamp is the Creation Date field in each saved notice's header,
which is a proxy for the publication time (Appendix~B.2). If a notice
counts as listed only from the end of its listing minute in AEMO's
public notice listing, 19.9\% of the matched window notices were listed
by the time they enter the forecasts and 80.1\% by 60 minutes after
that. Each main notice comparison is therefore repeated with
availability delayed by a further 5, 15 and 60 minutes (Appendix~B.3).

An article's saved version becomes available five minutes after the
later of its publication and modification times and is retained for 72
hours (Appendix~B.2). The documents available at issue time \(t\),
including those whose extraction failed, form the set \(\mathcal{D}_t\)
from which the text features of that forecast are computed.

\begin{figure}[tbp]
\centering
\includegraphics[width=\textwidth]{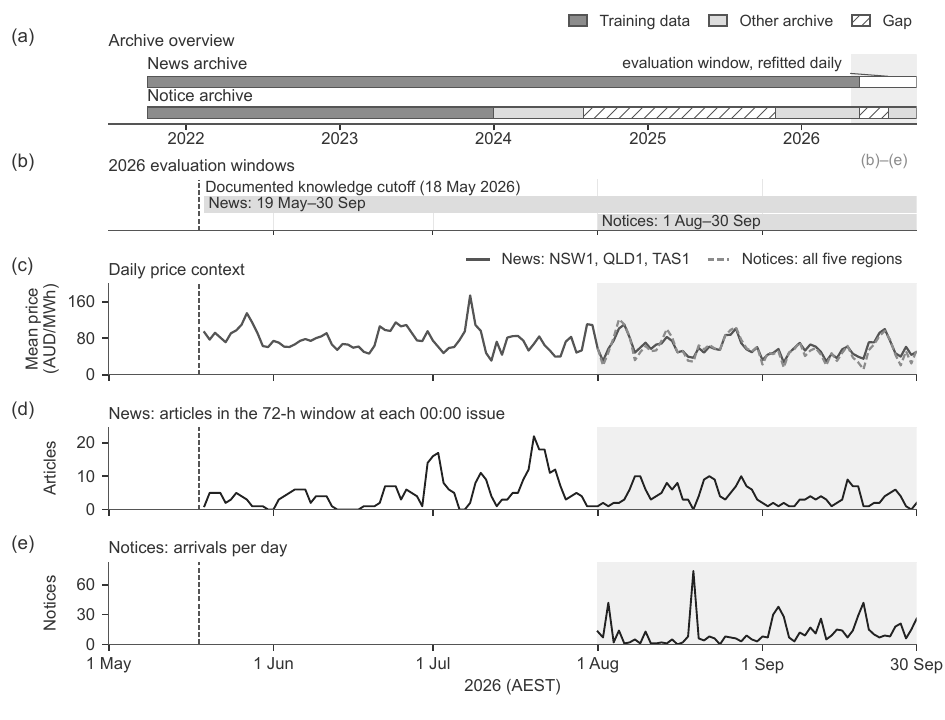}
\caption{Data archives, evaluation windows and daily context. (a) Saved archives
and training periods, 2021--2026; the shaded band marks the period of
(b)--(e). (b) Evaluation windows. (c) Daily mean target price. (d)
Articles in each 00:00 issue's 72-hour window. (e) Notice arrivals per
day. The dashed vertical line marks the extractor's documented knowledge
cutoff; shading in (c)--(e) marks the notice window.}\label{fig:2}
\end{figure}

\section{From text to features}\label{from-text-to-features}

Adding text to a model's numerical inputs supplies both document content
and the timing and volume of document arrivals. We compare three nested
models. The no-text model N uses the numerical inputs \(x_t\) available
at issue time \(t\). The activity reference M adds activity features
\(a(\mathcal{D}_t)\), which count the available documents and their
failed extractions, and a representation model L adds a representation
\(\varphi_L(\mathcal{D}_t)\) of their content:

\begin{equation}
N:\; x_t,\qquad M:\; \bigl(x_t,\ a(\mathcal{D}_t)\bigr),\qquad L:\; \bigl(x_t,\ a(\mathcal{D}_t),\ \varphi_L(\mathcal{D}_t)\bigr). \label{eq:1}
\end{equation}
\noindent The activity step compares M with N, the representation step compares
L with M, and the net change compares L with N. Each main comparison is
a representation step and measures the predictive increment of one
representation over the activity features in a fixed LightGBM
forecaster.

\subsection{Extraction}\label{extraction}

Both forecasting tasks take three elements from RE-Price
\citep{Chen2026REPrice}: the one-document extraction loop, the six
extraction attributes (impact timeframe, root cause, affected
infrastructure, region, energy resource and operational indicators) and
the sentence encoder that its benchmark models apply to the extracted
records. The news task also adopts its regions, availability rule for
edited articles, 72-hour input layout and impact tier; neither task uses
the RE-Price forecasting architecture. Each request contains only the
full title and body of one notice or article, with its identifier and
timestamps; the prompt tells the model to treat the source text as data,
to use no outside knowledge and to make no statements about prices or
later outcomes.

For a notice, the model records the reporting company and a lifecycle
state. For each impact window that the notice states, it records the
region, interval status, asset, fuel, event class and family, and the
six RE-Price attributes, and it binds each stated quantity to its unit,
role and window (Appendix~C.1). The windows follow
\citet{Lazarczyk2016}, who separates the time an operational message is
published from the interval of the event it reports. Evidence quotes and
free-text fields, such as stated times and quantities, are copied
verbatim; enumerated fields, such as region codes and event classes, are
classifications that this evidence must support. Deterministic code
normalises the times, and unstated fields stay UNKNOWN. A notice
forecasting a level-2 lack-of-reserve (LOR2) condition in Tasmania, for
example, yields a NEW record with a bounded TAS1 window from 09:00 to
11:00 and the 177 MW required and 164 MW available bound to it.

For an article, the model assigns the three-level RE-Price impact tier
(market-disruptive event, operational signal, or policy, regulatory or
survey item) and the same six attributes. It also extracts up to six
operational statements and quotes each verbatim; following Time-MMD
\citep{Liu2024aTimeMMD}, it classifies each as a sourced fact, a
forecast or an analysis.

We call the extractor, gpt-6-luna at medium reasoning effort, through
OpenAI's command-line client in a fresh session for
each request, with tools and web search disabled and a fixed JSON output
schema. The client exposed no temperature, seed, snapshot or cutoff
setting, so the serving snapshot and its actual knowledge cutoff cannot
be verified. An offline validator accepts a record only if every
evidence span occurs literally in the source and the windows and
quantities are structurally consistent. It checks quote provenance and
structure; it does not check whether the extracted content is correct.
In a human review of extracted records, 17 of 20 sampled articles
contained at least one unsupported extracted element, and 5 of 10
assessable notice impact windows were wrong (Appendix~C.3).

\subsection{Activity features and LLM
representations}\label{activity-features-and-llm-representations}

The activity features and the representation are computed on the same
available documents; failed extractions contribute to the activity
counts and enter the representation as unknown content. The activity
reference adds only document and failure counts; the failure counts come
from the LLM run and can depend on content.

For notices, the activity reference adds three NEM-wide counts, defined
in Eq.~(B.1) and repeated for every forecast region: the notices in the
saved history up to \(t\), the arrivals in the past six hours and the
cumulative number of failed extractions. In the training sample, the two
cumulative counts effectively form a time index; in the evaluation
window, they lie above their training range, so a tree learner treats
all their values alike. A windowed reference replaces the two cumulative
counts with the notices that became available in the past seven days and
the failed extractions among them, keeping the six-hour arrivals; a
version with 24-hour windows is a sensitivity analysis. Both use the
availability clock of Eq.~(B.1).

The notice representations describe the archive available at the issue
time through event states relevant to delivery (the forecast's target
interval), operational messages that overlap delivery and a cumulative
embedding of the extracted records. The Event states model adds 14
event-state columns, defined relative to the forecast region (events
there and in the other regions): active events by group, reserve
extremes and timing summaries. In these columns, a bounded record counts
only while its window contains the end of the target interval, an open-ended record only
during the 48 hours after it becomes available, and reserve quantities only
when bound to one window (Supplement~B.2).

The Overlap components model adds eight components after
\citet{Lazarczyk2016}, computed only for the forecast region. They count
the messages that overlap delivery and, among them, the
operational-failure records, in total and by fuel group. They also sum
the explicitly bound lost megawatts, which remain missing when a failure
record lacks a usable binding.

The Archive embeddings model adds an embedding of the cumulative archive
of structured records since October 2021, in which failed extractions
appear as UNKNOWN fields. The records are serialised in order of
availability, encoded in blocks of at most 254 tokens with the frozen
all-MiniLM-L6-v2 encoder of the Sentence-Transformers library
\citep{ReimersGurevych2019}, which returns 384-dimensional vectors, and
averaged over blocks (Appendix~C.2). Only 725 of the 21,470 records
(3.4\%) are added during the evaluation window, so the vector changes
slowly.

For news, the activity reference adds the counts of articles that became
available in the previous 72 hours and of their failed extractions. The
Fields model adds 34 fixed columns counting the extracted records: 4
impact-tier, 15 event-type, 4 statement-kind, 5 region and 6 target-date
counts. The Embeddings model adds an embedding of the extracted text:
each record's impact tier and six attributes are concatenated in a fixed
order and encoded with the same MiniLM encoder.

\subsection{Non-LLM representations and
placebos}\label{non-llm-representations-and-placebos}

Non-LLM representations test whether simpler readings of the same
available documents also improve on the activity reference. They read
the source text of every available document, including those whose LLM
extraction failed, and are added to the same activity reference. For
notices, the Rule parser model replaces the LLM records with rule-parser
records and projects them onto the same 14 event-state columns. The
notice Word frequencies model adds 24-hour word frequencies. For news,
the Raw embeddings model encodes the raw titles and bodies of the
available articles with the same MiniLM encoder, after the
headline-embedding benchmark of \citet{LopezLiraTang2026}. The news Word
frequencies model adds title and body word frequencies in the style of
\citet{Bai2024NewsLoad}, with a vocabulary screened on training data
(Supplement~C.2).

A time-shifted placebo rebuilds the whole text input, both the activity
features and the representation, from the documents available 28 days
earlier, whose source clocks and event dates move forward by 28 days.
Each placebo is refitted with the shifted text input in both training
and prediction, while its numerical inputs and targets stay unchanged.
The shifted text remains realistic but is misaligned with the target, as
in the negative-control exposures of \citet{Lipsitch2010NegCtrl}. In
both tasks, we rebuild the activity reference and the embedding model
this way (Archive embeddings for notices and Embeddings for news). The
notice placebos can be evaluated on the 38 target days from 24 August,
and the P5MIN extension includes a placebo of the same kind (Appendix~D.2).

\section{Forecasting design}\label{forecasting-design}

\subsection{Models}\label{models}

All LightGBM models of a task (Table~3) share one fixed configuration,
inherited from earlier development work and not tuned in this study
(\citealp{Ke2017LightGBM}; Appendix~B.4); only the input columns differ.
Each model is fitted to all regions and horizons of its task, which
enter as inputs. Notice models minimise absolute error, predict the
change from the latest available price and are fitted once on an
event-stratified sample from October 2021--December 2023 with sampling
weights (Appendix~B.4). News models minimise squared error and are
refitted every day on all earlier data from October 2021 (Appendix~B.1).

The no-text model, the activity references and the LLM representation
models are fitted at seeds 20260919, 20260920 and 20260921, and the
seed-averaged forecast of a model averages its three forecasts before
losses are computed. The non-LLM representations, the placebos and
LightGBM + P5MIN use seed 20260919, and Sections~6.2 and 6.3 report all
comparisons at that seed.

Two models without text provide benchmarks. For notices, the LightGBM +
P5MIN model gives the same learner eight input columns from AEMO's P5MIN
forecast; Appendix~D.2 reports an extension with notice text added to
this model. For news, a price-only version of LEAR \citep{Lago2021} fits
a LASSO regression for each hour to asinh-transformed prices
\citep{Uniejewski2018VST} of the same region one, two, three and seven
days earlier, with day-of-week indicators. As in \citet{Lago2021}, it is
recalibrated daily on four calibration windows whose forecasts are
averaged (Appendix~B.1).

\begin{table}[!htbp]
\begingroup\footnotesize
\setlength{\tabcolsep}{3pt}
\renewcommand{\arraystretch}{1.04}
\caption{Models and comparisons. Each model adds the listed inputs to the model
named first in its Inputs entry; N, M and L denote the nested models of
Eq.~(1) (--: outside the nesting).}\label{tab:3}
\begin{tabular}{>{\raggedright\arraybackslash}p{\dimexpr 0.23\linewidth-2\tabcolsep\relax}>{\raggedright\arraybackslash}p{\dimexpr 0.075\linewidth-2\tabcolsep\relax}>{\raggedright\arraybackslash}p{\dimexpr 0.385\linewidth-2\tabcolsep\relax}>{\raggedright\arraybackslash}p{\dimexpr 0.31\linewidth-2\tabcolsep\relax}}
\toprule
Model & In Eq.~(1) & Inputs & Comparison \\
\midrule
\addlinespace[0.6ex]
\multicolumn{4}{l}{\emph{Notices}}\\*
No-text model & N & Price history and anchor; physical market variables; calendar, region
and horizon & Base \\
\addlinespace[0.35ex]
Activity reference & M & No-text model + saved-history count, six-hour arrivals and cumulative
failed extractions & Reference \\
\addlinespace[0.35ex]
Windowed activity reference & M & No-text model + six-hour arrivals, seven-day arrivals and seven-day
failed extractions (sensitivity: 24-hour windows) & Reference (Section~6.1) \\
\addlinespace[0.35ex]
Event states & L & Activity reference + 14 event-state columns & Main (Holm over three comparisons) \\
\addlinespace[0.35ex]
Overlap components & L & Activity reference + eight overlap components & Main \\
\addlinespace[0.35ex]
Archive embeddings & L & Activity reference + embedding of the cumulative archive of structured
records & Main \\
\addlinespace[0.35ex]
Rule parser & L & Activity reference + rule-parser records on the 14 event-state columns & Non-LLM representation \\
\addlinespace[0.35ex]
Word frequencies & L & Activity reference + 24-hour word frequencies & Non-LLM representation \\
\addlinespace[0.35ex]
Placebos & M, L & Activity reference and Archive embeddings rebuilt from the notices
available 28 days earlier & Time-shifted placebo for Archive embeddings (38 target days) \\
\addlinespace[0.35ex]
LightGBM + P5MIN & -- & No-text model + P5MIN (eight columns) & Benchmark \\
\addlinespace[0.35ex]
P5MIN extension & -- & LightGBM + P5MIN + notice activity columns, then + 14 event-state
columns; time-shifted placebo of the latter & Extension (Appendix~D.2) \\
\addlinespace[0.35ex]
\addlinespace[0.6ex]
\multicolumn{4}{l}{\emph{News}}\\*
No-text model & N & Price history; physical market variables; calendar, region and horizon & Base \\
\addlinespace[0.35ex]
Activity reference & M & No-text model + numbers of available articles (72 h) and failed
extractions & Reference \\
\addlinespace[0.35ex]
Fields & L & Activity reference + 34 extracted count columns & Main (Holm over two comparisons) \\
\addlinespace[0.35ex]
Embeddings & L & Activity reference + embedding of the extracted text & Main \\
\addlinespace[0.35ex]
Raw embeddings & L & Activity reference + raw-text embedding & Non-LLM representation \\
\addlinespace[0.35ex]
Word frequencies & L & Activity reference + title and body word frequencies (57 and 81 words) & Non-LLM representation \\
\addlinespace[0.35ex]
Placebos & M, L & Activity reference and Embeddings rebuilt from the articles available 28
days earlier & Time-shifted placebo for Embeddings \\
\addlinespace[0.35ex]
LEAR & -- & Prices of the same region 1, 2, 3 and 7 days earlier; hourly LASSO,
recalibrated daily & Benchmark \\*
\bottomrule
\multicolumn{4}{p{\dimexpr\linewidth-2\tabcolsep\relax}}{\footnotesize\raggedright\setlength{\parindent}{0pt} \emph{Notes.} Each LLM notice representation is also added to the
windowed activity reference. Other comparisons, without Holm adjustment:
the activity reference and each LLM representation against the no-text
model (both tasks); LEAR against the no-text model, the activity
reference and the LLM representations (news).\par}\\
\end{tabular}\endgroup
\end{table}

\subsection{Loss and tests}\label{loss-and-tests}

Let \(m_{A,d}\) be the MAE of model \(A\) over the forecasts of day
\(d\), with unit weights on the complete grid, and \(m_A\) its MAE over
the window. For model \(A\), reference \(B\) and \(n\) days,

\begin{equation}
\Delta_{AB} = 100\,\frac{m_A - m_B}{m_B},\qquad
d_{AB,d} = m_{A,d} - m_{B,d},\qquad
\mathrm{DM}_{AB} = \frac{\sqrt{n}\,\bar d_{AB}}{s_{AB}}, \label{eq:2}
\end{equation}
\noindent where \(\bar d_{AB}\) is the mean of the daily differences and
\(s_{AB}\) their sample standard deviation, with divisor \(n - 1\) and
without autocovariance terms. Negative values favour \(A\). Effects are
reported as \(\Delta_{AB}\) and, in AUD/MWh, as \(\bar d_{AB}\). Using
the day as the unit of inference keeps the many forecasts within each
day from inflating the sample size, and the HAC7 and block-bootstrap
sensitivities below allow for dependence between days.

Each main comparison (Table~3) is tested by a two-sided daily DM test of
the fitted models' forecasts, using a standard normal reference
distribution
\citep{DieboldMariano1995,GiacominiWhite2006,Diebold2015,Lago2021}.
\citeauthor{Holm1979}'s (\citeyear{Holm1979}) procedure controls the
familywise error rate at 5\% within each task, and the tasks are tested
separately. In two sensitivities, HAC7 replaces \(s_{AB}^2\) by a
Bartlett heteroskedasticity- and autocorrelation-consistent (HAC)
variance with seven lags \citep{NeweyWest1987}, and the Student-t
sensitivity uses a t distribution with \(n - 1\) degrees of freedom
\citep{Harvey1997}, 60 for notices and 134 for news.

Paired, non-circular block bootstraps \citep{Kunsch1989} with one-day
and seven-day blocks and 100,000 draws give percentile intervals for
\(\Delta_{AB}\) that, like the tests, omit training randomness.
Bootstrap blocks of \(b\) days start uniformly at any of the
\(n - b + 1\) positions inside the window, without wrapping, and are
concatenated and truncated to \(n\) days; each draw recomputes the
reference MAE. Table~A1 reports these tests and intervals for the main
comparisons.

\section{Results}\label{results}

Table~4 and Figure~3 report the activity step, the representation step
and the net change with seed-averaged forecasts; Tables~5 and 6 report
all comparisons at seed 20260919, including the non-LLM representations,
the placebos and the benchmarks. Significance refers to two-sided daily
DM tests at the 5\% level, Holm-adjusted for the main comparisons and
nominal otherwise. LightGBM + P5MIN lowers notice MAE by 3.23\% relative
to the no-text model, and LEAR lowers news MAE by 22.14\%; all text
models built on the no-text model have a higher MAE than these
benchmarks.

\subsection{Activity and representation steps with seed-averaged
forecasts}\label{activity-and-representation-steps-with-seed-averaged-forecasts}

\begin{table}[!htbp]
\begingroup\footnotesize
\setlength{\tabcolsep}{3pt}
\renewcommand{\arraystretch}{1.04}
\caption{Activity step, representation step and net change with seed-averaged
forecasts. \(\Delta\)MAE (\%) of Eq.~(2), with the nominal two-sided DM \emph{p}
in parentheses; negative values favour the model with more inputs.}\label{tab:4}
\begin{tabular}{>{\raggedright\arraybackslash}p{\dimexpr 0.22\linewidth-2\tabcolsep\relax}>{\raggedright\arraybackslash}p{\dimexpr 0.17\linewidth-2\tabcolsep\relax}>{\raggedleft\arraybackslash}p{\dimexpr 0.17\linewidth-2\tabcolsep\relax}>{\raggedleft\arraybackslash}p{\dimexpr 0.19\linewidth-2\tabcolsep\relax}>{\raggedleft\arraybackslash}p{\dimexpr 0.17\linewidth-2\tabcolsep\relax}>{\raggedright\arraybackslash}p{\dimexpr 0.08\linewidth-2\tabcolsep\relax}}
\toprule
Representation & Activity reference & \mbox{Activity step}\newline\mbox{(M vs N)} & \mbox{Representation step}\newline\mbox{(L vs M)} & \mbox{Net change}\newline\mbox{(L vs N)} & Signs differ \\
\midrule
\addlinespace[0.6ex]
\multicolumn{6}{l}{\emph{Notices}}\\*
Event states & Original & +0.224 (0.050) & −0.143 (\textless0.001) & +0.081 (0.506) & Yes \\
\addlinespace[0.35ex]
 & Seven-day window & +0.081 (0.003) & −0.141 (\textless0.001) & −0.060 (0.050) & No \\
\addlinespace[0.35ex]
 & 24-hour window & +0.107 (\textless0.001) & −0.212 (\textless0.001) & −0.106 (\textless0.001) & No \\
\addlinespace[0.35ex]
Overlap components & Original & +0.224 (0.050) & −0.004 (0.838) & +0.219 (0.060) & Yes \\
\addlinespace[0.35ex]
 & Seven-day window & +0.081 (0.003) & +0.047 (0.074) & +0.128 (\textless0.001) & No \\
\addlinespace[0.35ex]
 & 24-hour window & +0.107 (\textless0.001) & +0.017 (0.540) & +0.124 (\textless0.001) & No \\
\addlinespace[0.35ex]
Archive embeddings & Original & +0.224 (0.050) & −0.517 (\textless0.001) & −0.295 (0.029) & No \\
\addlinespace[0.35ex]
 & Seven-day window & +0.081 (0.003) & −0.521 (0.002) & −0.440 (0.014) & No \\
\addlinespace[0.35ex]
 & 24-hour window & +0.107 (\textless0.001) & −0.448 (\textless0.001) & −0.342 (0.005) & No \\
\addlinespace[0.35ex]
\addlinespace[0.6ex]
\multicolumn{6}{l}{\emph{News}}\\*
Fields & 72-hour counts & −3.03 (0.004) & +12.50 (0.049) & +9.09 (0.148) & No \\
\addlinespace[0.35ex]
Embeddings & 72-hour counts & −3.03 (0.004) & −0.53 (0.757) & −3.55 (0.049) & No \\*
\bottomrule
\multicolumn{6}{p{\dimexpr\linewidth-2\tabcolsep\relax}}{\footnotesize\raggedright\setlength{\parindent}{0pt} \emph{Notes.} Seed-averaged forecasts average the forecasts of each
model at seeds 20260919, 20260920 and 20260921 before losses are
computed; notices: 61 target days; news: 135 issue days. Original: the
activity references of Table~3. The windowed notice references replace
the two cumulative counts with counts over the past seven days or 24
hours (Section~4.2). Signs differ: the net change and the representation
step have opposite signs. Holm-adjusted \emph{p} for the three
representation steps against the seven-day reference: \textless0.001
(Event states), 0.074 (Overlap components) and 0.004 (Archive
embeddings). With seed-averaged forecasts, the no-text model has an MAE
of 9.4142 AUD/MWh for notices and 24.695 AUD/MWh for news. Seven-day
block-bootstrap intervals are in Supplement~D.1.\par}\\
\end{tabular}\endgroup
\end{table}

\begin{figure}[!htbp]
\centering
\includegraphics[width=\textwidth]{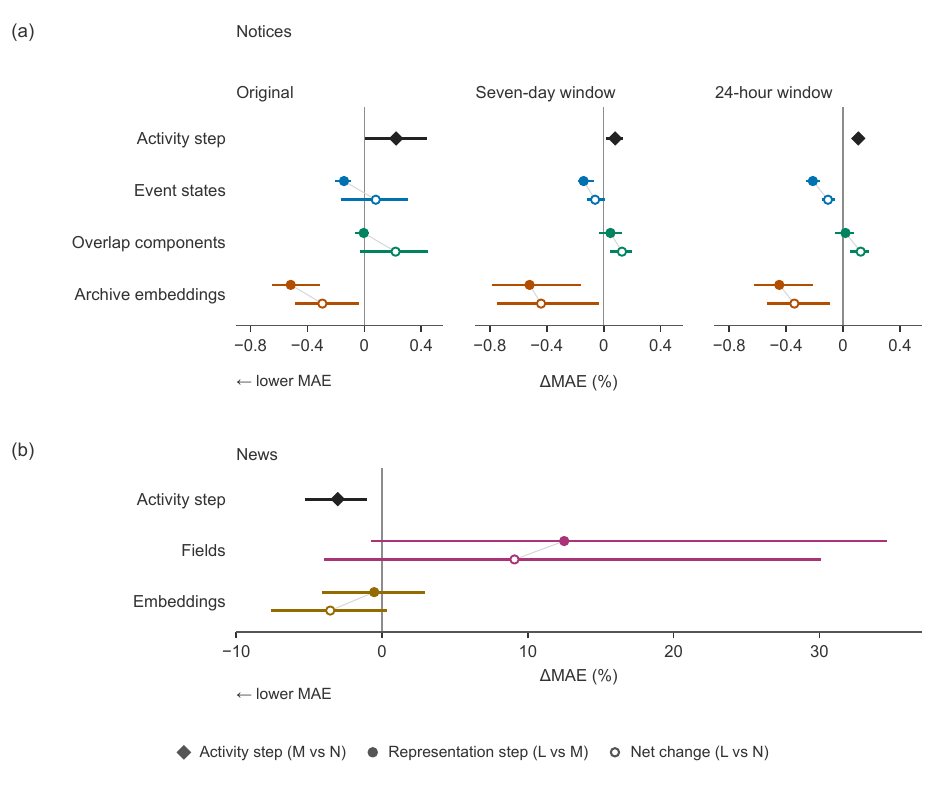}
\caption{Activity step, representation step and net change. With seed-averaged
forecasts, Event states and Archive embeddings lower MAE beyond activity
under all three notice references, and for news the activity step makes
up most of the net change of the Embeddings model. Each point is
\(\Delta_{AB}\) of Eq.~(2) for seed-averaged forecasts; negative values
favour the model with more inputs. Diamonds show the activity step (M vs
N) of each activity reference, filled circles the representation step (L
vs M) and open circles the net change (L vs N). Lines are 95\% seven-day
block-bootstrap intervals, which condition on the fitted forecasts. (a)
Notices, 61 target days, with the original, seven-day and 24-hour
references; (b) news, 135 issue days, on a different scale. Values are
in Table~4 and the intervals in Supplement~D.1.}\label{fig:3}
\end{figure}

For news, the activity step makes up most of the net change of the
Embeddings model (Table~4, Figure~3(b)). The article and
failed-extraction counts lower MAE by 3.03\% (\mbox{\emph{p} = 0.004};
seven-day block interval −5.28\% to −1.04\%). The Embeddings model
lowers MAE by 3.55\% relative to the no-text model (\mbox{\emph{p} = 0.049};
seven-day block interval −7.60\% to +0.34\%) and by 0.53\% relative to
the activity reference (\mbox{\emph{p} = 0.757}). The Fields model raises MAE
by 12.50\% relative to the activity reference (nominal \mbox{\emph{p} = 0.049}).

Two of the three LLM notice representations add small gains beyond
activity under every notice reference (Table~4, Figure~3(a)). Against
the original reference, Event states and Archive embeddings lower MAE by
0.143\% and 0.517\% (both \mbox{\emph{p} \textless{} 0.001}), and Overlap
components leaves it unchanged (−0.004\%, \mbox{\emph{p} = 0.838}). Against the
seven-day reference, the reductions are 0.141\% and 0.521\% (\mbox{Holm \emph{p} \textless{} 0.001} and 0.004), and Overlap components raises MAE
by 0.047\% (\mbox{\emph{p} = 0.074}); the 24-hour reference gives reductions of
0.212\% and 0.448\% and a change of +0.017\%. In AUD/MWh, the reductions
of Event states and Archive embeddings are 0.01--0.05.

The notice activity step shrinks when the cumulative counts are removed.
The original notice counts raise MAE by 0.224\% (\mbox{\emph{p} = 0.050}), the
seven-day counts by 0.081\% (\mbox{\emph{p} = 0.003}) and the 24-hour counts by
0.107\% (\mbox{\emph{p} \textless{} 0.001}). At seed 20260919, the original
counts raise MAE by 0.546\% (Table~5) and the seven-day counts by
0.234\%.

Gains against the no-text model and against the activity reference
differ in sign for some representations, mostly at individual seeds
(Table~D2). With the original notice reference and the news reference,
such reversals occur for three representations at seed 20260919 (Event
states, Overlap components and news Embeddings), two at seed 20260920
(Archive embeddings and news Embeddings) and one at seed 20260921
(Overlap components), and for Event states and Overlap components in the
seed-averaged forecasts. With the windowed notice references, they occur
at one seed each and not in the seed-averaged forecasts, and the news
Embeddings model shows none in its seed-averaged forecasts.

\subsection{Notices at seed 20260919}\label{notices-at-seed-20260919}

Relative to the activity reference (Table~5), the three LLM notice
representations lower MAE by 0.30\% (Event states), 0.28\% (Overlap
components) and 0.75\% (Archive embeddings). In AUD/MWh, these
reductions are 0.03--0.07, compared with 0.30 from adding the P5MIN
inputs to the no-text model (Table~5). Each reduction is significant
after Holm adjustment, also under HAC7 and the Student-t distribution,
and all block-bootstrap intervals exclude zero (Table~A1).

Of the three LLM representations, only the Archive embeddings model has
a lower MAE than the no-text model, by 0.205\%. The activity features
raise MAE by 0.55\%; the Event states and Overlap components models
recover about half of this increase but remain 0.24\% and 0.26\% above
the no-text model. Against the no-text model, the nominal DM p-values
are 0.037, 0.012 and 0.041 for Event states, Overlap components and
Archive embeddings, and all three seven-day block-bootstrap intervals
include zero (Supplement~D.1).

\needspace{3\baselineskip}
Whether the Archive embeddings model is more accurate than its
time-shifted placebo depends on the activity reference. On the 38
placebo days and with the original reference, it has a 0.745\% lower MAE
than its placebo. The time-shifted Archive embeddings model has a 0.39\%
higher MAE than the time-shifted activity reference, and both
differences are significant (Table~5). With the seven-day reference, the
Archive embeddings model has a 0.04\% higher MAE than its placebo
(\mbox{\emph{p} = 0.726}). In a descriptive comparison on these days, the
original, unshifted activity reference has a 0.18\% higher MAE than its
time-shifted placebo (Appendix~D.2).

The non-LLM Rule parser and Word frequencies models also lower MAE
relative to the activity reference, by 0.29\% and 0.41\%, and both
reductions are significant; these point estimates lie within the range
of the LLM representations. In direct comparisons, Archive embeddings
has a 0.45\% lower MAE than Rule parser and a 0.34\% lower MAE than Word
frequencies (both \mbox{\emph{p} \textless{} 0.001}). Event states and Overlap
components do not differ significantly from Rule parser and have 0.11\%
and 0.13\% higher MAEs than Word frequencies (\mbox{\emph{p} = 0.018} and
0.021).

In the P5MIN extension (Table~D3), adding the notice activity features
and then Event states to the LightGBM + P5MIN benchmark lowers its MAE
by 0.79\% in total and by 0.16\% in the last step, and both reductions
are significant. On the 38 placebo days, however, the resulting model
has a significantly higher MAE than its time-shifted placebo, by 0.26\%.

\begin{table}[!htbp]
\begingroup\footnotesize
\setlength{\tabcolsep}{1.5pt}
\renewcommand{\arraystretch}{1.04}
\caption{Notice task: MAE and paired comparisons (five-minute prices 5 to 30
minutes ahead, five regions; 61 target days, 1 August--30 September
2026; 527,040 forecasts). MAE in AUD/MWh.}\label{tab:5}
\begin{tabular}{>{\raggedright\arraybackslash}p{\dimexpr 0.33\linewidth-2\tabcolsep\relax}>{\raggedright\arraybackslash}p{\dimexpr 0.12\linewidth-2\tabcolsep\relax}>{\raggedleft\arraybackslash}p{\dimexpr 0.1\linewidth-2\tabcolsep\relax}>{\raggedleft\arraybackslash}p{\dimexpr 0.115\linewidth-2\tabcolsep\relax}>{\raggedleft\arraybackslash}p{\dimexpr 0.155\linewidth-2\tabcolsep\relax}>{\raggedleft\arraybackslash}p{\dimexpr 0.09\linewidth-2\tabcolsep\relax}>{\raggedleft\arraybackslash}p{\dimexpr 0.09\linewidth-2\tabcolsep\relax}}
\toprule
\multiheader{Model} & \multiheader{Compared with} & \multiheader{MAE} & \multicolumn{2}{c}{\(\Delta\)MAE} & \multicolumn{2}{c}{\textit{p}} \\
\cmidrule(lr){4-5}\cmidrule(lr){6-7}
 & & & \% & AUD/MWh & DM & Holm \\
\midrule
\mbox{No-text model} & \mbox{--} & \mbox{9.4093} & \mbox{--} & \mbox{--} & \mbox{--} & \mbox{--} \\
\addlinespace[0.35ex]
\addlinespace[0.6ex]
\multicolumn{7}{l}{\emph{Activity step}}\\*
\mbox{Activity reference} & \mbox{N} & \mbox{9.4606} & \mbox{+0.546} & \mbox{+0.051} & \mbox{\textless0.001} & \mbox{--} \\
\addlinespace[0.35ex]
\addlinespace[0.6ex]
\multicolumn{7}{l}{\emph{Representation step: LLM representations}}\\*
\mbox{Event states} & \mbox{M} & \mbox{9.4320} & \mbox{−0.303} & \mbox{−0.029} & \mbox{\textless0.001} & \mbox{\textless0.001} \\
\addlinespace[0.35ex]
\mbox{Overlap components} & \mbox{M} & \mbox{9.4337} & \mbox{−0.284} & \mbox{−0.027} & \mbox{\textless0.001} & \mbox{\textless0.001} \\
\addlinespace[0.35ex]
\mbox{Archive embeddings} & \mbox{M} & \mbox{9.3900} & \mbox{−0.747} & \mbox{−0.071} & \mbox{\textless0.001} & \mbox{\textless0.001} \\
\addlinespace[0.35ex]
\addlinespace[0.6ex]
\multicolumn{7}{l}{\emph{Representation step: non-LLM representations}}\\*
\mbox{Rule parser} & \mbox{M} & \mbox{9.4328} & \mbox{−0.294} & \mbox{−0.028} & \mbox{\textless0.001} & \mbox{--} \\
\addlinespace[0.35ex]
\mbox{Word frequencies} & \mbox{M} & \mbox{9.4216} & \mbox{−0.413} & \mbox{−0.039} & \mbox{\textless0.001} & \mbox{--} \\
\addlinespace[0.35ex]
\addlinespace[0.6ex]
\multicolumn{7}{l}{\emph{Net change: LLM representations}}\\*
\mbox{Event states} & \mbox{N} & \mbox{9.4320} & \mbox{+0.242} & \mbox{+0.023} & \mbox{0.037} & \mbox{--} \\
\addlinespace[0.35ex]
\mbox{Overlap components} & \mbox{N} & \mbox{9.4337} & \mbox{+0.260} & \mbox{+0.024} & \mbox{0.012} & \mbox{--} \\
\addlinespace[0.35ex]
\mbox{Archive embeddings} & \mbox{N} & \mbox{9.3900} & \mbox{−0.205} & \mbox{−0.019} & \mbox{0.041} & \mbox{--} \\
\addlinespace[0.35ex]
\addlinespace[0.6ex]
\multicolumn{7}{l}{\emph{Time-shifted placebo (38 target days, 24 August--30 September)}}\\*
\mbox{Archive embeddings} & \mbox{Placebo} & \mbox{9.5192} & \mbox{−0.745} & \mbox{−0.071} & \mbox{\textless0.001} & \mbox{--} \\
\addlinespace[0.35ex]
\mbox{Archive embeddings,}\newline\mbox{time-shifted} & \mbox{Shifted M} & \mbox{9.5906} & \mbox{+0.390} & \mbox{+0.037} & \mbox{\textless0.001} & \mbox{--} \\
\addlinespace[0.35ex]
\addlinespace[0.6ex]
\multicolumn{7}{l}{\emph{Benchmark with operator-forecast inputs}}\\*
\mbox{LightGBM + P5MIN} & \mbox{N} & \mbox{9.1051} & \mbox{−3.233} & \mbox{−0.304} & \mbox{\textless0.001} & \mbox{--} \\*
\bottomrule
\multicolumn{7}{p{\dimexpr\linewidth-2\tabcolsep\relax}}{\footnotesize\raggedright\setlength{\parindent}{0pt} \emph{Notes.} \(\Delta\)MAE (\%): \(\Delta_{AB}\) of Eq.~(2), negative values
favouring the model in the first column; \(\Delta\)MAE (AUD/MWh): mean daily
difference. DM \emph{p}: nominal two-sided daily DM test. Holm \emph{p}:
adjusted within the three main notice comparisons. All comparisons use
seed 20260919; Tables~4 and D1 give the seed-averaged forecasts and the
other seeds. Time-shifted placebo rows: Archive embeddings and the
activity reference rebuilt from the notices available 28 days earlier;
the time-shifted activity reference has an MAE of 9.5533 on these days.
--: not applicable. Reference abbreviations: N, no-text model; M,
activity reference; Placebo, time-shifted placebo; Shifted M, shifted
activity reference.\par}\\
\end{tabular}\endgroup
\end{table}

\subsection{News at seed 20260919}\label{news-at-seed-20260919}

Neither LLM news representation improves on the activity reference
(Table~6). The Fields and Embeddings models have 11.41\% and 1.17\%
higher MAEs, and neither difference is significant after Holm
adjustment. The seven-day block interval for Embeddings ranges from a
3.11\% reduction to a 5.94\% increase (Table~A1).

Adding the article and failed-extraction counts to the no-text model
lowers MAE by 3.58\%; the reduction is significant, also under HAC7, and
both of its block-bootstrap intervals exclude zero (Table~6). The
time-shifted activity reference, rebuilt from the articles available 28
days earlier, does not improve on the no-text model (+1.52\%, \mbox{\emph{p} = 0.280}). Relative to the no-text model, the Embeddings model has a 2.45\%
lower MAE and the Fields model a 7.43\% higher MAE, but neither
difference is significant.

Among the four news representations, the Embeddings model has the lowest
MAE, but none improves on the activity reference: the Raw embeddings and
Word frequencies models raise MAE by 12.04\% and 21.15\%. In direct
comparisons, the Embeddings model has a 9.70\% lower MAE than the Raw
embeddings model (\mbox{\emph{p} = 0.071}) and a 16.49\% lower MAE than the
Word frequencies model (\mbox{\emph{p} = 0.046}). The Embeddings model does not
differ significantly from its time-shifted placebo.

\begin{table}[!htbp]
\begingroup\footnotesize
\setlength{\tabcolsep}{1.5pt}
\renewcommand{\arraystretch}{1.04}
\caption{News task: MAE and paired comparisons (day-ahead hourly prices for NSW1,
QLD1 and TAS1, issued at 00:00 AEST; 135 issue days, 19 May--30
September 2026; 9,720 forecasts). MAE in AUD/MWh.}\label{tab:6}
\begin{tabular}{>{\raggedright\arraybackslash}p{\dimexpr 0.33\linewidth-2\tabcolsep\relax}>{\raggedright\arraybackslash}p{\dimexpr 0.12\linewidth-2\tabcolsep\relax}>{\raggedleft\arraybackslash}p{\dimexpr 0.1\linewidth-2\tabcolsep\relax}>{\raggedleft\arraybackslash}p{\dimexpr 0.115\linewidth-2\tabcolsep\relax}>{\raggedleft\arraybackslash}p{\dimexpr 0.155\linewidth-2\tabcolsep\relax}>{\raggedleft\arraybackslash}p{\dimexpr 0.09\linewidth-2\tabcolsep\relax}>{\raggedleft\arraybackslash}p{\dimexpr 0.09\linewidth-2\tabcolsep\relax}}
\toprule
\multiheader{Model} & \multiheader{Compared with} & \multiheader{MAE} & \multicolumn{2}{c}{\(\Delta\)MAE} & \multicolumn{2}{c}{\textit{p}} \\
\cmidrule(lr){4-5}\cmidrule(lr){6-7}
 & & & \% & AUD/MWh & DM & Holm \\
\midrule
\mbox{No-text model} & \mbox{--} & \mbox{24.813} & \mbox{--} & \mbox{--} & \mbox{--} & \mbox{--} \\
\addlinespace[0.35ex]
\addlinespace[0.6ex]
\multicolumn{7}{l}{\emph{Activity step}}\\*
\mbox{Activity reference} & \mbox{N} & \mbox{23.925} & \mbox{−3.58} & \mbox{−0.89} & \mbox{0.017} & \mbox{--} \\
\addlinespace[0.35ex]
\addlinespace[0.6ex]
\multicolumn{7}{l}{\emph{Representation step: LLM representations}}\\*
\mbox{Fields} & \mbox{M} & \mbox{26.656} & \mbox{+11.41} & \mbox{+2.73} & \mbox{0.043} & \mbox{0.086} \\
\addlinespace[0.35ex]
\mbox{Embeddings} & \mbox{M} & \mbox{24.205} & \mbox{+1.17} & \mbox{+0.28} & \mbox{0.558} & \mbox{0.558} \\
\addlinespace[0.35ex]
\addlinespace[0.6ex]
\multicolumn{7}{l}{\emph{Representation step: non-LLM representations}}\\*
\mbox{Raw embeddings} & \mbox{M} & \mbox{26.805} & \mbox{+12.04} & \mbox{+2.88} & \mbox{0.041} & \mbox{--} \\
\addlinespace[0.35ex]
\mbox{Word frequencies} & \mbox{M} & \mbox{28.986} & \mbox{+21.15} & \mbox{+5.06} & \mbox{0.035} & \mbox{--} \\
\addlinespace[0.35ex]
\addlinespace[0.6ex]
\multicolumn{7}{l}{\emph{Net change: LLM representations}}\\*
\mbox{Fields} & \mbox{N} & \mbox{26.656} & \mbox{+7.43} & \mbox{+1.84} & \mbox{0.204} & \mbox{--} \\
\addlinespace[0.35ex]
\mbox{Embeddings} & \mbox{N} & \mbox{24.205} & \mbox{−2.45} & \mbox{−0.61} & \mbox{0.305} & \mbox{--} \\
\addlinespace[0.35ex]
\addlinespace[0.6ex]
\multicolumn{7}{l}{\emph{Time-shifted placebos}}\\*
\mbox{Time-shifted}\newline\mbox{activity reference} & \mbox{N} & \mbox{25.190} & \mbox{+1.52} & \mbox{+0.38} & \mbox{0.280} & \mbox{--} \\
\addlinespace[0.35ex]
\mbox{Embeddings} & \mbox{Placebo} & \mbox{24.205} & \mbox{−4.37} & \mbox{−1.10} & \mbox{0.253} & \mbox{--} \\
\addlinespace[0.35ex]
\addlinespace[0.6ex]
\multicolumn{7}{l}{\emph{Price-only benchmark}}\\*
\mbox{LEAR} & \mbox{N} & \mbox{19.319} & \mbox{−22.14} & \mbox{−5.49} & \mbox{\textless0.001} & \mbox{--} \\
\addlinespace[0.35ex]
\mbox{LEAR} & \mbox{M} & \mbox{19.319} & \mbox{−19.25} & \mbox{−4.61} & \mbox{\textless0.001} & \mbox{--} \\*
\bottomrule
\multicolumn{7}{p{\dimexpr\linewidth-2\tabcolsep\relax}}{\footnotesize\raggedright\setlength{\parindent}{0pt} \emph{Notes.} As for Table~5, with Holm \emph{p} adjusted within the two
main news comparisons; the time-shifted placebo rows rebuild the
activity reference and Embeddings from the articles available 28 days
earlier, and the time-shifted placebo of Embeddings has an MAE of
25.310. For the activity step, the HAC7 \emph{p} is 0.025, and the 95\%
block-bootstrap intervals are −6.43\% to −0.90\% (one-day blocks) and
−6.57\% to −0.80\% (seven-day blocks). Reference abbreviations: N,
no-text model; M, activity reference; Placebo, time-shifted placebo.\par}\\
\end{tabular}\endgroup
\end{table}

\subsection{Sensitivity analyses}\label{sensitivity-analyses}

The notice reductions accumulate across the window (Figure~4(a)) and
persist after excluding the first seven target days (Table~D1). Delaying
availability by 5, 15 or 60 minutes changes these reductions by less
than 0.01 percentage points (Appendix~B.3).

\begin{figure}[!hbp]
\centering
\includegraphics[width=\textwidth]{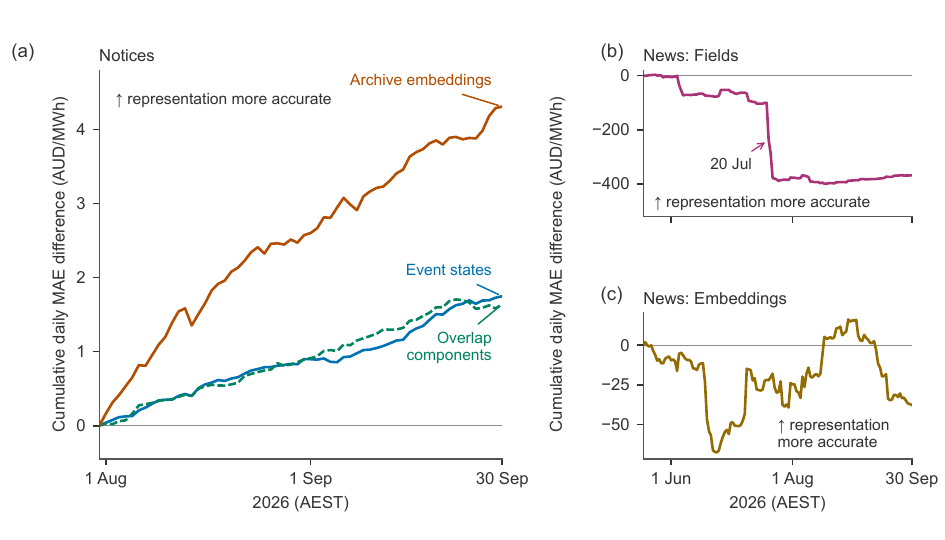}
\caption{Cumulative daily MAE differences. The notice reductions accumulate
across the window, whereas the news Fields loss concentrates around 20
July. Each curve is the cumulative sum of the activity reference's daily
MAE minus the representation model's (AUD/MWh, seed 20260919), as in
\citet{WelchGoyal2008}; a rising curve means the representation model is
more accurate. (a) Notices, 61 target days; (b) news Fields and (c) news
Embeddings, 135 issue days, on separate scales. The annotation in (b)
marks 20 July 2026.}\label{fig:4}
\end{figure}

Changing the evaluation loss or the activity reference gives a different
pattern. Under squared error, only the Overlap components reduction is
significant (Table~D1). Against an alternative activity reference that
also receives the notice type from the source metadata, the Event states
model lowers MAE by 0.13\% and the Archive embeddings model by 0.48\%,
whereas the Overlap components model raises it by 0.44\%; all three
differences are significant after Holm adjustment.

The representation steps also vary with the training seed (Table~D1).
Against the original reference, seed 20260919 gives the largest notice
reductions; with seed 20260920, no representation lowers MAE
significantly, and the Overlap components model is significantly worse
than its activity reference, by 0.30\%. With the seven-day reference,
none of the three reductions at seed 20260919 is significant (Supplement~D.1).

For news, both cumulative curves end below zero (Figures~4(b) and 4(c)),
and no sensitivity analysis shows a significant improvement (Table~D1).
The largest single-day change for Fields occurs on 20 July, when
dispatch prices stay below 140 AUD/MWh in every region and the daily MAE
of the Fields model reaches 150 AUD/MWh against 12.8 AUD/MWh for the
activity reference. On that day, the 72-hour window holds more articles
than on any other day of the news window (Figure~2(d)).

\begingroup\spaceskip=0.29em plus 0.20em minus 0.12em\tolerance=2000
\section{Discussion}\label{discussion}

\looseness=-1 In the news task, the article and failed-extraction counts alone achieve
3.03 of the 3.55 percentage points by which the Embeddings model
improves on the no-text model, and the content adds no significant gain
beyond them. Published text gains for NEM prices are measured against
no-text models (Table~1) and leave this activity step unmeasured. As in
macroeconomic forecasting, where simple text metrics add less as the
numerical benchmark becomes richer \citep{Kalamara2022}, the news text
gain shrinks once the reference contains the document counts.

\looseness=-1 How the notices are counted changes the notice activity step. With
seed-averaged forecasts, the original counts, two of which act as a time
index, raise MAE by 0.224\%, and counts over a recent window raise it by
0.081--0.107\%; the representation steps of Event states and Archive
embeddings are reductions of 0.14--0.21\% and 0.45--0.52\% under all
three references. A reversal occurs when the activity and representation
steps have opposite signs and the activity step has the larger absolute
value in AUD/MWh. For notices, both steps are a few tenths of a percent,
so the training seed can decide which is larger.

\looseness=-1 With seed-averaged forecasts and the seven-day reference, the Archive
embeddings model is more accurate than both the no-text model (−0.44\%)
and the activity reference (−0.52\%). The time-shifted placebos give
mixed evidence on whether such gains depend on content aligned with the
target. At seed 20260919, the Archive embeddings model has a lower MAE
than its time-shifted placebo with the original reference (−0.745\%) and
a higher MAE with the seven-day reference (+0.04\%); the archive
embedding changes slowly and could act partly as a time index, and
embeddings of recent notices alone were not tested. At the same seed,
adding the notice activity features and then Event states to the
stronger LightGBM + P5MIN benchmark lowers MAE at each step (Section~6.2), and the resulting model is less accurate than its time-shifted
placebo.

\looseness=-1 The activity features lower news MAE and raise notice MAE. Besides the
text source, the two tasks differ in horizon, timing rule, training
objective, refitting schedule and activity window. The news reference
counts articles over the past 72 hours, whereas the original notice
reference combines six-hour arrivals with cumulative counts (Section~4.2). The notice activity step also depends on the base model: added to
LightGBM + P5MIN, the original notice activity features enlarge its MAE
reduction relative to the no-text model from 3.23\% to 3.84\% (Table~D3).

\looseness=-1 The evidence covers one market and one news source over windows of 61
and 135 days. Notice release times are approximated by a header
timestamp. The tested delays extend to 60 minutes, whereas the latest
matched listing comes 714 minutes after the notice enters the forecasts
(Appendix~B.2). The numerical inputs are final stored values available
under the timing rules of Appendix~B.1, because revision histories are
unavailable. The results are therefore retrospective comparisons under
these timing rules, and real-time gains are not verified.

\looseness=-1 The notice models were fitted once on data ending in December 2023, more
than 2.5 years before evaluation, and the news models minimise squared
error but are evaluated by MAE. One fixed LightGBM configuration
produces all text models of a task.

\looseness=-1 How the extraction errors found in the human review (Section~4.1) affect
forecast loss is not estimated. The training data also include documents
that predate the extractor's documented knowledge cutoff, and hindsight
in their extractions could affect the relations that the forecaster
learns \citep{LopezLiraTangZhu2025}. The serving model's actual cutoff
is unverified, and the size and direction of this effect are unknown.

\section{Conclusion}\label{conclusion}

\looseness=-1 This paper proposes an activity reference for evaluating LLM-derived
text features, which adds counts of the same available documents and of
their failed extractions to the no-text model and needs no additional
LLM requests. We apply it to price forecasts for Australia's National
Electricity Market issued after the extractor's documented knowledge
cutoff, averaging the forecasts of the no-text model, the activity
references and the LLM representation models over three training seeds.
The reference splits each gain over the no-text model into an activity
step and a representation step. For WattClarity news, the counts lower
MAE by 3.0\%, most of the 3.6\% gain of the better LLM representation,
and neither LLM representation improves significantly on the activity
reference.

\looseness=-1 For AEMO market notices, Event states and Archive embeddings lower MAE
beyond the activity features by 0.14--0.21\% and 0.45--0.52\% with the
original and the windowed references, and Overlap components shows no
significant reduction. At a single seed, rule-parser and word-frequency
representations of the same notices, which need no LLM, give reductions
within the range of the LLM representations. At that seed and with the
seven-day reference, the Archive embeddings model is no more accurate
than its time-shifted placebo. In both tasks, the benchmark is more
accurate than every text model built on the no-text model.

\looseness=-1 Because document counts alone can achieve much of a text gain measured
only against a model without text, text-feature evaluations should
report the activity step and the representation step. They should also
count documents over a recent window, average forecasts over training
seeds and include time-shifted placebos where they can be constructed.
Further tests would use prospective windows with versioned extractor
snapshots, real-time listing clocks and stronger base learners.

\par\endgroup
\begingroup\small
\section*{Data and code availability}\label{data-and-code-availability}

AEMO market data and market notices are public. The code and data are
publicly available at
\mbox{\url{https://github.com/ningfanxing-sys/nem-llm-text-activity}},
including the configuration files, saved
forecasts, daily losses and the notice extraction records. WattClarity
articles are copyrighted and are not redistributed; the repository
provides their article identifiers and derived numeric features. The
extraction prompts, the JSON schemas, a feature dictionary
(feature\_dictionary.md) and the supplementary details
(supplementary\_details.md, cited as Supplement~A.1--D.2) are also in the
repository.

\section*{Declaration of generative AI and AI-assisted technologies in the writing process}

During the preparation of this work, the authors used Claude and ChatGPT
to assist with writing, coding and editing. After using these tools, the
authors reviewed and edited the content as needed and take full
responsibility for the content of the publication.

\endgroup

\appendix
\numberwithin{equation}{section}
\section{Alternative tests and
intervals}\label{alternative-tests-and-intervals}

Table~A1 reports the HAC7 and Student-t versions of the daily DM test
and the block-bootstrap intervals of Section~5.2 for the main
comparisons.

\setcounter{table}{0}\renewcommand{\thetable}{A\arabic{table}}\renewcommand{\theHtable}{A\arabic{table}}
\begin{table}[!htbp]
\begingroup\footnotesize
\setlength{\tabcolsep}{3pt}
\renewcommand{\arraystretch}{1.04}
\caption{Alternative tests and 95\% bootstrap intervals for the main comparisons
(seed 20260919).}\label{tab:A1}
\begin{tabular}{>{\raggedright\arraybackslash}p{\dimexpr 0.4\linewidth-2\tabcolsep\relax}>{\raggedright\arraybackslash}p{\dimexpr 0.19999999999999998\linewidth-2\tabcolsep\relax}>{\raggedright\arraybackslash}p{\dimexpr 0.19999999999999998\linewidth-2\tabcolsep\relax}>{\raggedright\arraybackslash}p{\dimexpr 0.19999999999999998\linewidth-2\tabcolsep\relax}}
\toprule
\multicolumn{4}{l}{\textbf{Panel A. Notices}}\\
Item & Event states & Overlap components & Archive embeddings \\
\midrule
HAC7 \emph{p} & \textless0.001 & \textless0.001 & \textless0.001 \\*
Student-t \emph{p} & \textless0.001 & \textless0.001 & \textless0.001 \\*
95\% interval, one-day blocks & {[}−0.391, −0.218{]} & {[}−0.379, −0.188{]} & {[}−0.977, −0.514{]} \\*
95\% interval, seven-day blocks & {[}−0.401, −0.220{]} & {[}−0.380, −0.193{]} & {[}−0.926, −0.486{]} \\*
\end{tabular}
\par\nointerlineskip
\begin{tabular}{>{\raggedright\arraybackslash}p{\dimexpr 0.4\linewidth-2\tabcolsep\relax}>{\raggedright\arraybackslash}p{\dimexpr 0.3\linewidth-2\tabcolsep\relax}>{\raggedright\arraybackslash}p{\dimexpr 0.3\linewidth-2\tabcolsep\relax}}
\midrule
\multicolumn{3}{l}{\textbf{Panel B. News}}\\
Item & Fields (news) & Embeddings (news) \\
\midrule
HAC7 \emph{p} & 0.159 & 0.604 \\*
Student-t \emph{p} & 0.045 & 0.559 \\*
95\% interval, one-day blocks & {[}1.99, 24.05{]} & {[}−2.68, 5.15{]} \\*
95\% interval, seven-day blocks & {[}−0.10, 31.36{]} & {[}−3.11, 5.94{]} \\*
\bottomrule
\end{tabular}
\par\smallskip{\footnotesize\raggedright\setlength{\parindent}{0pt} \emph{Notes.} Intervals: paired percentile block-bootstrap intervals for
the percentage change in MAE relative to the activity reference (Eq.~(2); Tables~5 and 6). HAC7 and Student-t \emph{p}-values are nominal and
two-sided.\par}
\endgroup
\end{table}

\FloatBarrier
\section{Information timing and learner
settings}\label{information-timing-and-learner-settings}

\subsection{Numerical inputs, news training labels and
LEAR}\label{numerical-inputs-news-training-labels-and-lear}

A numerical record enters a forecast issued at time \(t\) only once it
is available under its own source clock: a price at the later of its
interval end plus five minutes and its last-change time (LASTCHANGED),
the other dispatch-summary inputs five minutes after their interval end,
and a P5MIN run once its run time plus five minutes and its last-change
time are both no later than \(t\). Stored values are final, because
revision histories are not available. The latest available five-minute
price is the anchor, and the news task never uses the price interval
ending at 00:00. At each 00:00 issue \(t\), the news models are trained
on every complete target day from 1 October 2021 whose 72 hourly labels
were all available strictly before \(t\). LEAR selects the LASSO penalty
using the Akaike information criterion and averages with equal weights
the forecasts from four calibration windows of 56, 84, 1,092 and 1,456
days \citep{Lago2021}. Supplement~B.1 gives the full input list, the
news preprocessing and the LEAR implementation.

\subsection{Document availability and listing
evidence}\label{document-availability-and-listing-evidence}

Because article availability follows the saved version (Section~3.3), an
edit can make an older article available at a later issue; through such
edits and the 72-hour lookback, 37 of the 207 articles available in the
news window were first published before the window began. Notices in the
training sample enter at their official clock plus five minutes, rounded
up to the five-minute grid. Evaluation-period notices enter at
\(a_0 = \lceil C + 5\ \text{min} \rceil_5\), where the Creation Date
\(C\) is a field in the header of the saved notice text that does not
establish the notice's first publication time. Extraction latency is not
modelled: a record becomes available together with its document. With
\(a_j\) the availability time of notice \(j\) and \(F\) the set of
notices whose extraction failed, the notice activity reference adds,
over all regions and all saved notices since the start of the archive,

\begin{equation}
P(t) = \#\{j : a_j \le t\},\quad A_6(t) = \#\{j : t - 6\,\text{h} < a_j \le t\},\quad P_F(t) = \#\{j \in F : a_j \le t\}. \label{eq:B1}
\end{equation}

Updates, cancellations and recovery notices close earlier records only
after they themselves become available, so no later notice revises an
earlier information set (Supplement~B.2). Of the 725 notices dated
August or September 2026, 712 match a record in AEMO's public notice
listing; listing evidence covers 142 of them (19.9\%) at \(a_0\) and 570
(80.1\%) after 60 minutes, and the latest listing occurs 714 minutes
after \(a_0\).

\subsection{Delayed availability and date
parsing}\label{delayed-availability-and-date-parsing}

The delayed conditions make every evaluation-window notice available 5,
15 or 60 minutes after \(a_0\) and rebuild every notice-derived input
from the notices available by then; the models fitted with seed 20260919
predict without refitting. The main estimates read the DD/MM/YYYY
timestamps inside the 796 evaluation-period notices day first, as AEMO
writes them, and a sensitivity analysis reads them month first. The two
readings differ only in the Event states, Overlap components and Rule
parser features of the evaluation window, so the same fitted models
produce both sets of forecasts (Table~D1; Supplement~B.3).

\subsection{Training sample and LightGBM
settings}\label{training-sample-and-lightgbm-settings}

The notice training sample keeps every training example in connected
high-price events, built from five-minute prices above 300 AUD/MWh, with
weight 1, and a one-in-50 hash sample of the others, with weight 50
(Supplement~B.4). Table~B1 lists the settings, inherited from earlier
development configurations and used without tuning, early stopping or
seed selection in this study; both tasks use LightGBM 4.6.0
\citep{Ke2017LightGBM}.

\begingroup\setlength{\intextsep}{0pt}
\setcounter{table}{0}\renewcommand{\thetable}{B\arabic{table}}\renewcommand{\theHtable}{B\arabic{table}}
\begin{table}[!hbp]
\begingroup\footnotesize\captionsetup{skip=4pt}
\setlength{\tabcolsep}{3pt}
\renewcommand{\arraystretch}{1.04}
\caption{LightGBM settings.}\label{tab:B1}
\begin{tabular}{>{\raggedright\arraybackslash}p{\dimexpr 0.26\linewidth-2\tabcolsep\relax}>{\raggedright\arraybackslash}p{\dimexpr 0.37\linewidth-2\tabcolsep\relax}>{\raggedright\arraybackslash}p{\dimexpr 0.37\linewidth-2\tabcolsep\relax}}
\toprule
Setting & Notices & News \\
\midrule
Objective & Absolute error & Squared error \\
\addlinespace[0.35ex]
Training sample & October 2021--December 2023 sample, fitted once; target is the change
from the latest available price; sampling weights 1 and 50 & All earlier complete days from 1 October 2021, refitted daily \\
\addlinespace[0.35ex]
Learning rate & 0.04 & 0.04 \\
\addlinespace[0.35ex]
Trees & 180, no early stopping & 180, no early stopping \\
\addlinespace[0.35ex]
Leaves; maximum depth & 15; 5 & 15; 5 \\
\addlinespace[0.35ex]
Minimum examples per leaf & 100 & 100 \\
\addlinespace[0.35ex]
Column fraction per tree & 0.85 & 0.85 \\
\addlinespace[0.35ex]
Row bagging & None & None \\
\addlinespace[0.35ex]
L1; L2 penalty & 0; 1 & 0; 1 \\
\addlinespace[0.35ex]
Deterministic training & Yes & Yes \\
\addlinespace[0.35ex]
Prediction clipping & −1,000 to 17,500 AUD/MWh & None \\
\bottomrule
\end{tabular}\endgroup
\end{table}
\endgroup

\FloatBarrier
\section{Extraction and human review}\label{extraction-and-human-review}

\subsection{Prompts, schemas and extraction
runs}\label{prompts-schemas-and-extraction-runs}

The verbatim prompts and JSON schemas are in the code repository; Supplement~C.1 lists the schema fields and gives a worked example.

Notices were extracted in one run and articles in two, a historical run
(September 2021--February 2026) and a 2026 run. All three runs used the
same model alias and reasoning effort, and the two article runs also
used the same prompt and schema. Extraction failed for 297 of the 11,348
notices in the training sample (2.62\%) and 13 of the 725 dated August
or September 2026 (1.79\%), and for 201 of the 2,244 articles in the
historical run (8.96\%) and 26 of the 262 articles from March to
September in the 2026 run (9.92\%). The historical article run retried
documents whose outputs failed validation, whereas the 2026 run kept its
validation failures as final. Supplement~C.1 gives the run times, the
retry rules and the remaining counts.

\subsection{Encoding}\label{encoding}

Supplement~C.2 gives the encoder version, the serialisation and the
vocabulary screening of the news Word frequencies model, and the
file feature\_dictionary.md in the code repository defines every model column.

\subsection{Human review}\label{human-review}

We reviewed month-stratified samples of successful extractions
from 2026: 20 news articles with 66 statements from March--September
2026, and 20 notices, 10 each from August and September 2026. The
samples are not weighted to the corpus. We assigned every code. In the news sample, 17 of the
20 articles contained at least one extracted element without support in
the source (Wilson 95\% interval 0.64--0.95), and 8 had omissions; the
event time was wrong in 23 of the 65 statements for which it could be
assessed. Table~C1 gives the notice results.

\setcounter{table}{0}\renewcommand{\thetable}{C\arabic{table}}\renewcommand{\theHtable}{C\arabic{table}}
\begin{table}[!hbp]
\begingroup\footnotesize
\setlength{\tabcolsep}{3pt}
\renewcommand{\arraystretch}{1.04}
\caption{Human review of notice records: errors in our review of 20
notices and agreement with a second annotator on 5 of them.}\label{tab:C1}
\begin{tabular}{>{\raggedright\arraybackslash}p{\dimexpr 0.4\linewidth-2\tabcolsep\relax}>{\raggedleft\arraybackslash}p{\dimexpr 0.3\linewidth-2\tabcolsep\relax}>{\raggedleft\arraybackslash}p{\dimexpr 0.3\linewidth-2\tabcolsep\relax}}
\toprule
Field & Errors / assessable & Second annotator: agreement \\
\midrule
Region & 1/12 & 5/5 \\
\addlinespace[0.35ex]
Event type & 3/11 & 5/5 \\
\addlinespace[0.35ex]
Lifecycle state & 1/12 & 2/5 \\
\addlinespace[0.35ex]
Impact window & 5/10 & 2/5 \\
\addlinespace[0.35ex]
Quantity binding & 1/12 & 4/5 \\
\addlinespace[0.35ex]
Notice linkage & not assessable (0/0) & 5/5 (all not applicable) \\
\addlinespace[0.35ex]
Completeness & 3/18 & 2/5 \\
\addlinespace[0.35ex]
Overlap-component projection & 0/10 & 5/5 \\
\addlinespace[0.35ex]
Total over the first seven fields & 14/75 & -- \\*
\bottomrule
\multicolumn{3}{p{\dimexpr\linewidth-2\tabcolsep\relax}}{\footnotesize\raggedright\setlength{\parindent}{0pt} \emph{Notes.} Errors are counted over assessable judgements; fields
within a notice are correlated, so the seven-field total has no
interval. Agreement uses the raw codes, including unknown and
not-applicable codes; for linkage, both annotators coded every item as
not applicable. Agreement is reported by field only (--: not
applicable).\par}\\
\end{tabular}\endgroup
\end{table}

A second annotator known to us coded five of the notices and
five of the articles without seeing our answers; both annotators knew that the labels were LLM outputs. For
the five articles, the two codings agreed on unsupported elements in
four and on omissions and the impact tier in all five; across the eight
statements of these articles, they agreed on the magnitude in six and on
every other statement field in all eight.

\FloatBarrier
\section{Additional results}\label{additional-results}

\subsection{Seeds and sensitivity
analyses}\label{seeds-and-sensitivity-analyses}

For each seed, the no-text model, the activity reference and every LLM
representation model are refitted with that seed, and the seed-averaged
forecast averages each model's three predictions before losses are
computed (Table~D1). For news, no seed gives a significant improvement
over the activity reference for either representation, and the
Embeddings estimate changes sign across seeds. Table~D2 gives the
activity steps and the sign reversals at each seed.

\setcounter{table}{0}\renewcommand{\thetable}{D\arabic{table}}\renewcommand{\theHtable}{D\arabic{table}}
\begin{table}[!b]
\begingroup\footnotesize
\setlength{\tabcolsep}{1.5pt}
\renewcommand{\arraystretch}{1.35}
\caption{Sensitivity analyses of the main comparisons: change in MAE (\(\Delta\), \%)
relative to the activity reference unless stated (negative values favour
the representation), and nominal two-sided DM \emph{p} unless the row
states Holm.}\label{tab:D1}
\begin{tabular}{>{\raggedright\arraybackslash}p{\dimexpr 0.31\linewidth-2\tabcolsep\relax}>{\raggedleft\arraybackslash}p{\dimexpr 0.11499999999999999\linewidth-2\tabcolsep\relax}>{\raggedleft\arraybackslash}p{\dimexpr 0.11499999999999999\linewidth-2\tabcolsep\relax}>{\raggedleft\arraybackslash}p{\dimexpr 0.11499999999999999\linewidth-2\tabcolsep\relax}>{\raggedleft\arraybackslash}p{\dimexpr 0.11499999999999999\linewidth-2\tabcolsep\relax}>{\raggedleft\arraybackslash}p{\dimexpr 0.11499999999999999\linewidth-2\tabcolsep\relax}>{\raggedleft\arraybackslash}p{\dimexpr 0.11499999999999999\linewidth-2\tabcolsep\relax}}
\toprule
\multicolumn{7}{l}{\textbf{Panel A. Notices}}\\
Variant & \multicolumn{2}{c}{Event states} & \multicolumn{2}{c}{Overlap components} & \multicolumn{2}{c}{Archive embeddings} \\
 & \(\Delta\) (\%) & \textit{p} & \(\Delta\) (\%) & \textit{p} & \(\Delta\) (\%) & \textit{p} \\
\midrule
Seed 20260919 & −0.303 & \textless0.001 & −0.284 & \textless0.001 & −0.747 & \textless0.001 \\*
Seed 20260920 & +0.066 & 0.136 & +0.299 & \textless0.001 & −0.027 & 0.785 \\*
Seed 20260921 & −0.214 & \textless0.001 & −0.017 & 0.705 & −0.492 & \textless0.001 \\*
Seed-averaged forecast & −0.143 & \textless0.001 & −0.004 & 0.838 & −0.517 & \textless0.001 \\*
Squared error & −0.157 & 0.252 & −0.597 & \textless0.001 & −0.029 & 0.876 \\*
First seven notice target days excluded (54 days) & −0.298 & \textless0.001 & −0.268 & \textless0.001 & −0.694 & \textless0.001 \\*
Alternative activity reference with notice-type metadata (Holm~$p$) & −0.129 & 0.001 & +0.442 & \textless0.001 & −0.476 & \textless0.001 \\*
Day-first date parsing (main estimates); availability at \(a_0\) or 5,
15 or 60 min later (Holm~$p$) & −0.309\newline to −0.303 & \textless0.001 & −0.285\newline to −0.284 & \textless0.001 & −0.750\newline to −0.746 & \textless0.001 \\*
Month-first date parsing; availability at \(a_0\) or 5, 15 or 60 min
later (Holm~$p$) & −0.309\newline to −0.303 & \textless0.001 & −0.285\newline to −0.284 & \textless0.001 & −0.750\newline to −0.746 & \textless0.001 \\*
\end{tabular}
\par\nointerlineskip
\begin{tabular}{>{\raggedright\arraybackslash}p{\dimexpr 0.31\linewidth-2\tabcolsep\relax}>{\raggedleft\arraybackslash}p{\dimexpr 0.1725\linewidth-2\tabcolsep\relax}>{\raggedleft\arraybackslash}p{\dimexpr 0.1725\linewidth-2\tabcolsep\relax}>{\raggedleft\arraybackslash}p{\dimexpr 0.1725\linewidth-2\tabcolsep\relax}>{\raggedleft\arraybackslash}p{\dimexpr 0.1725\linewidth-2\tabcolsep\relax}}
\midrule
\multicolumn{5}{l}{\textbf{Panel B. News}}\\
Variant & \multicolumn{2}{c}{Fields} & \multicolumn{2}{c}{Embeddings (news)} \\
 & \(\Delta\) (\%) & \textit{p} & \(\Delta\) (\%) & \textit{p} \\
\midrule
Seed 20260919 & +11.41 & 0.043 & +1.17 & 0.558 \\*
Seed 20260920 & +16.10 & 0.040 & +0.43 & 0.801 \\*
Seed 20260921 & +9.80 & 0.093 & −1.61 & 0.465 \\*
Seed-averaged forecast & +12.50 & 0.049 & −0.53 & 0.757 \\*
Squared error & +26.22 & 0.109 & −3.05 & 0.538 \\*
Extended news window, 1 March--30 September 2026 (214 days) & +8.46 & 0.035 & +2.14 & 0.168 \\*
\bottomrule
\end{tabular}
\par\smallskip{\footnotesize\raggedright\setlength{\parindent}{0pt} \emph{Notes.} Squared-error entries are percentage changes in mean
squared error. The alternative activity reference also receives the
notice type from the source metadata. Date-parsing rows give ranges
across the availability conditions of Appendix~B.3.\par}
\endgroup
\end{table}
\FloatBarrier
\setcounter{table}{1}\renewcommand{\thetable}{D\arabic{table}}\renewcommand{\theHtable}{D\arabic{table}}
\begin{table}[!t]
\begingroup\footnotesize
\setlength{\tabcolsep}{3pt}
\renewcommand{\arraystretch}{1.04}
\caption{Activity steps and sign reversals by seed. Activity step: \(\Delta\)MAE (\%) of
the activity reference against the no-text model, with the nominal DM
\emph{p} in parentheses. Reversals: LLM representations whose net change
and representation step have opposite signs.}\label{tab:D2}
\begin{tabular}{>{\raggedright\arraybackslash}p{\dimexpr 0.2\linewidth-2\tabcolsep\relax}>{\raggedleft\arraybackslash}p{\dimexpr 0.2\linewidth-2\tabcolsep\relax}>{\raggedleft\arraybackslash}p{\dimexpr 0.2\linewidth-2\tabcolsep\relax}>{\raggedleft\arraybackslash}p{\dimexpr 0.2\linewidth-2\tabcolsep\relax}>{\raggedleft\arraybackslash}p{\dimexpr 0.2\linewidth-2\tabcolsep\relax}}
\toprule
 & Seed 20260919 & Seed 20260920 & Seed 20260921 & Seed-averaged \\
\midrule
\addlinespace[0.6ex]
\multicolumn{5}{l}{\emph{Notices, original reference}}\\*
Activity step & +0.546 (\textless0.001) & +0.109 (0.348) & +0.065 (0.604) & +0.224 (0.050) \\
\addlinespace[0.35ex]
Reversals & Event states, Overlap components & Archive embeddings & Overlap components & Event states, Overlap components \\
\addlinespace[0.35ex]
\addlinespace[0.6ex]
\multicolumn{5}{l}{\emph{Notices, seven-day window}}\\*
Activity step & +0.234 (\textless0.001) & +0.056 (0.210) & −0.009 (0.852) & +0.081 (0.003) \\
\addlinespace[0.35ex]
Reversals & Event states, Overlap components, Archive embeddings & None & None & None \\
\addlinespace[0.35ex]
\addlinespace[0.6ex]
\multicolumn{5}{l}{\emph{Notices, 24-hour window}}\\*
Activity step & +0.195 (\textless0.001) & +0.440 (\textless0.001) & −0.266 (\textless0.001) & +0.107 (\textless0.001) \\
\addlinespace[0.35ex]
Reversals & None & Event states, Overlap components, Archive embeddings & None & None \\
\addlinespace[0.35ex]
\addlinespace[0.6ex]
\multicolumn{5}{l}{\emph{News}}\\*
Activity step & −3.58 (0.017) & −4.45 (\textless0.001) & −1.24 (0.368) & −3.03 (0.004) \\
\addlinespace[0.35ex]
Reversals & Embeddings & Embeddings & None & None \\
\bottomrule
\end{tabular}\endgroup
\end{table}
\setcounter{table}{2}\renewcommand{\thetable}{D\arabic{table}}\renewcommand{\theHtable}{D\arabic{table}}
\begin{table}[!t]
\begingroup\footnotesize
\setlength{\tabcolsep}{3pt}
\renewcommand{\arraystretch}{1.04}
\caption{P5MIN extension (notices) and the news representations against LEAR.}\label{tab:D3}
\begin{tabular}{>{\raggedright\arraybackslash}p{\dimexpr 0.55\linewidth-2\tabcolsep\relax}>{\raggedleft\arraybackslash}p{\dimexpr 0.15\linewidth-2\tabcolsep\relax}>{\raggedleft\arraybackslash}p{\dimexpr 0.18\linewidth-2\tabcolsep\relax}>{\raggedleft\arraybackslash}p{\dimexpr 0.12\linewidth-2\tabcolsep\relax}}
\toprule
Comparison & \(\Delta\)MAE (\%) & \(\Delta\)MAE (AUD/MWh) & DM \emph{p} \\
\midrule
\addlinespace[0.6ex]
\multicolumn{4}{l}{\emph{Notices: P5MIN extension}}\\*
(LightGBM + P5MIN) vs no-text model & −3.233 & −0.304 & \textless0.001 \\
\addlinespace[0.35ex]
(LightGBM + P5MIN + activity) vs no-text model & −3.839 & −0.361 & \textless0.001 \\
\addlinespace[0.35ex]
(LightGBM + P5MIN + activity + Event states) vs no-text model & −3.993 & −0.376 & \textless0.001 \\
\addlinespace[0.35ex]
(LightGBM + P5MIN + activity + Event states) vs (LightGBM + P5MIN) & −0.785 & −0.072 & \textless0.001 \\
\addlinespace[0.35ex]
(LightGBM + P5MIN + activity + Event states) vs (LightGBM + P5MIN +
activity) & −0.160 & −0.014 & 0.002 \\
\addlinespace[0.35ex]
(LightGBM + P5MIN + activity + Event states) vs time-shifted placebo (38
days, 24 August--30 September) & +0.263 & +0.024 & \textless0.001 \\
\addlinespace[0.35ex]
\addlinespace[0.6ex]
\multicolumn{4}{l}{\emph{News: representations against LEAR}}\\*
Fields vs LEAR & +37.97 & +7.34 & \textless0.001 \\
\addlinespace[0.35ex]
Embeddings vs LEAR & +25.29 & +4.89 & \textless0.001 \\*
\bottomrule
\multicolumn{4}{p{\dimexpr\linewidth-2\tabcolsep\relax}}{\footnotesize\raggedright\setlength{\parindent}{0pt} \emph{Notes.} LightGBM + P5MIN: the no-text model with eight P5MIN
columns; + activity: the three notice activity columns; + Event states:
the 14 event-state columns; time-shifted placebo: the activity columns
and Event states rebuilt from the notices available 28 days earlier. The
three LLM notice representation models have MAEs 3.1\% to 3.6\% higher
than LightGBM + P5MIN (Table~5).\par}\\
\end{tabular}\endgroup
\end{table}
\subsection{Placebos, the P5MIN extension and
LEAR}\label{placebos-the-p5min-extension-and-lear}

On the 38 notice target days from 24 August, the Archive embeddings
model has a 0.745\% lower MAE than its time-shifted placebo (95\%
one-day block interval −0.998\% to −0.499\%; \mbox{\emph{p} \textless{} 0.001}), and the activity reference has a 0.184\% higher MAE than its
time-shifted placebo in a descriptive comparison. For news, the
difference between Embeddings and its time-shifted placebo is not
significant (\mbox{\emph{p} = 0.253}). In the other placebo comparisons, the
time-shifted notice activity reference does not differ significantly
from the no-text model (+0.157\%, \mbox{\emph{p} = 0.22}), the time-shifted
Archive embeddings model has a 0.547\% higher MAE than the no-text model
(\mbox{\emph{p} \textless{} 0.001}), and the time-shifted news Embeddings model
does not differ significantly from the time-shifted news activity
reference (+0.478\%, \mbox{\emph{p} = 0.90}). Supplement~D.2 gives the MAE
levels on the placebo days. Table~D3 reports the P5MIN extension,
including the time-shifted placebo of the LightGBM + P5MIN model with
Event states, and the news representations against LEAR.

\FloatBarrier

\newcommand{\doi}[1]{doi:\,\href{https://doi.org/#1}{#1}}
\begingroup
\small
\linespread{0.89}\selectfont
\raggedright
\setlength{\bibsep}{0pt}
\interlinepenalty=10000
\providecommand{\DOIprefix}{}
\bibliographystyle{elsarticle-harv}
\bibliography{refs/references}
\endgroup

\end{document}